\documentclass[12pt]{article}
\pdfoutput=1

\usepackage{amsmath, amsfonts, amssymb,dsfont,upgreek}
\usepackage{graphicx}
\usepackage{psfrag}
\usepackage[usenames,dvipsnames,svgnames,table]{xcolor}
\usepackage{enumerate}
\usepackage{jheppub}
\usepackage{soul}
\usepackage[utf8]{inputenc}
\usepackage{subfig}
\usepackage{float}

\newcommand{\be}{\begin{equation}}
\newcommand{\ee}{\end{equation}}
\newcommand{\vol}{\textrm{vol}}
\newcommand{\w}{\wedge}

\def\be{\begin{equation}}
\def\ee{\end{equation}}
\def\bea{\begin{eqnarray}}
\def\eea{\end{eqnarray}}

\newcommand{\ep}{\epsilon}

\renewcommand{\d}{\textrm{d}}
\newcommand{\e}{\textrm{e}}
\newcommand{\V}{\mathcal{V}}

\renewcommand{\(}{\left(}
\renewcommand{\)}{\right)}

\usepackage{marginnote}
\definecolor{onecolor}{rgb}{1, .83,.83}
\definecolor{twocolor}{rgb}{.83, .9,.83}

\def\simleq{\; \raise0.3ex\hbox{$<$\kern-0.75em
      \raise-1.1ex\hbox{$\sim$}}\; }
   \def\simgeq{\; \raise0.3ex\hbox{$>$\kern-0.75em
      \raise-1.1ex\hbox{$\sim$}}\; }
      
      \newcommand{\figref}[1]{Fig.\,\ref{#1}}

\newcommand{\secref}[1]{Sec.\,\ref{#1}}

\title{Towards an explicit model of large field inflation}
\author[\spadesuit]{Juan Diaz Dorronsoro}
\author{and}
\author[\heartsuit]{Marjorie Schillo}

\emailAdd{juan@itf.fys.kuleuven.be, marjorie.schillo@physics.uu.se}

\affiliation[\spadesuit]{\it Institute for Theoretical Physics, KU Leuven, 3001 Leuven, Belgium  }
\affiliation[\heartsuit]{\it Institutionen f\"or fysik och astronomi, Uppsala University,
Uppsala, Sweden  }

\begin{document}

\abstract{The unwinding inflation mechanism is studied in a type IIB flux compactification where all moduli are stabilized using flux, non-perturbative effects, and the leading $\alpha'$ corrections of the large volume scenario. We consider the backreaction on the geometry due to the presence of anti-D3 branes as well as the backreaction of inflation on the K\"ahler moduli, and compute the resulting corrections to the slow-roll potential.  By taking large flux numbers, we are able to find inflationary epochs where backreaction effects are under control, the inflaton traverses a super-Planckian field range, and the resulting amplitude of scalar perturbations is consistent with observation.  }

\maketitle

\section{Introduction}
An impressively large number of effective field theories (EFT) of inflation succeed in reproducing the observed properties of the spectrum of primordial fluctuations as measured by the Planck collaboration \cite{Ade:2015lrj}.  While this may be considered a triumph of the inflationary paradigm, in order to move forward with our exploration of the early universe and the high energy physics that was at play, it is necessary to cull inflationary models that cannot be embedded in a UV complete theory of quantum gravity (read: string theory). Since the  parameters of UV complete models of inflation should be fixed by the vacuum expectation values (vevs) of string  moduli, it is reasonable to expect that this set of models will be far more restricted, and therefore far more predictive, than the full set of EFT models.   A UV complete embedding of inflation is desirable for its increased predictivity and the glimpse it provides of quantum gravity, and it is moreover essential for large-field theories of inflation, where the inflaton traverses a super-Planckian field range and therefore the expansion in $\phi/\Lambda$ (field over cutoff) implicit in any EFT construction, does not converge. 

There is a growing literature of various ``swampland" conjectures \cite{Vafa:2005ui, Ooguri:2006in, ArkaniHamed:2006dz, Brennan:2017rbf} which aims at formulating general principles which can be used to distinguish UV-completable EFT's, that lie somewhere in the string landscape, from the rest which are relegated to the swampland. The importance of understanding such underlying principles of quantum gravity is difficult to overstate, however to date we lack a proof for any of the conjectures.  While there is a growing body of suggestive arguments stemming from AdS/CFT, black hole physics, and folk theorems, some of the primary evidence for the conjectures lies in a lack of counterexamples.  Thus, we view the best way to further support or sharpen said conjectures is to search for a counterexample.  Specifically, we aim to find a  large field model of inflation that takes into account backreaction on string moduli.  This would help to sharpen and quantify the claim of \cite{Ooguri:2006in}: ``We cannot have a slow roll inflation where the distance in the scalar moduli space is much bigger than Planck length and still use the same effective field
theory.''

We succeed in finding a large field model that takes into account the backreaction on string moduli by embedding the unwinding inflation mechanism \cite{D'Amico:2012ji, Gautason:2016cyp} in a Klebanov-Strassler \cite{Klebanov:2000hb} (KS) throat region of a compact manifold.  We use the setting of warped  orientifold compactifications of type IIB on  Calabi-Yau three-folds \cite{Giddings:2001yu} where all complex structure moduli are stabilized at tree level.  Additionally, we use the large volume scenario (LVS) \cite{Balasubramanian:2005zx,Conlon:2005ki} to break the no-scale structure and stabilize a model  with two K\"ahler moduli. The unwinding mechanism achieves slow roll inflation by gradually decreasing the positive vacuum energy sourced by anti-D3 branes. This is mediated by a 5-brane bubble which expands, crossing many times over the $S^3$ at the tip of the KS geometry.  As the 5-brane bubble moves across the $S^3$, it removes both three-form flux and anti-D3 branes via brane-flux annihilation \cite{Kachru:2002gs}.  The repeated motion over a compact cycle, removing flux with each pass , realizes a flux cascade \cite{Kleban:2011cs} which gradually decreases the four-dimensional vacuum energy. For a more detailed description of this process, see \secref{explainmechanism}.

One of the main purposes of this work is to fortify the original embedding of unwinding inflation in string theory \cite{Gautason:2016cyp} by taking into account backreaction.  The backreaction effects we consider are 1) the effect of anti-D3 brane charge on the warping of the KS geometry, and 2) the evolution of the K\"ahler moduli's potential due to the depletion of anti-D3 charge during the unwinding process.  We compute these effects and find scenarios in which they do not spoil inflation.  While we are able to explicitly compute the contribution to the slow roll potential due to  K\"ahler moduli backreaction, we only consider the leading UV effect of the backreacted antibranes.  An explicit calculation of backreaction involving the supersymmetry breaking effects of the antibranes remains beyond the scope of this work, however the backreaction of anti-D3 branes in the KS geometry is extremely well-studied \cite{DeWolfe:2008zy, Bena:2009xk,Bena:2011hz,Bena:2011wh,Dymarsky:2011pm,Dymarsky:2013tna}, making future efforts to compute these higher order backreaction particularly tractable.  Further, we estimate the effects of supersymmetry breaking backreaction and argue that it is tunably small in the UV. Due to this well-studied setting, this model serves as an ideal setting to further solidify or rule out the claim of a trans-Planckian field range.

We find that in order to obtain large field ranges where backreaction effects do not spoil inflation and all moduli are stabilized at high mass, we need large numbers of three-form flux in the KS throat.  Unless there are additional sources of negative D3 charge outside of the throat region, which enters with opposite sign in the tadpole condition \eqref{tadpole}, the three-from flux in the KS region implies an Euler characteristic for the associated Calabi-Yau four-fold that is larger than known examples. Consequently, we do not have an explicit global example of a Calabi-Yau three-fold which accommodates our model; hence \emph{towards} an explicit model of large field inflation. 

Beyond the consideration of underlying principles in quantum gravity, we are also interested in whether this model can reproduce the observed cosmic microwave background (CMB.)  After all, it is the aim of every model of inflation to explain the observed features in the CMB and additionally teach us about the high energy physics at play in the early universe.  In this respect, the current embedding of the model has passed the first test for remaining a viable model of inflation.  We are able to achieve an inflationary period of at least 60 efolds that gives rise to the correct amplitude of Gaussian curvature perturbations.  We further expect a rich phenomenology including equilateral and resonant non-Gaussianity and observable tensors. However, the details of these observables will be sensitive to the explicit details of the global Calabi-Yau three-fold through the highly non-linear constraints that Calabi-Yau with different moduli masses will impose on parameter space.  In light of this, we see little point in computing specific observables in lieu of a fully explicit embedding.  Rather, we take the embedding using a generic Swiss-cheese style manifold with two K\"ahler moduli as a proof of principle that this model is viable and warrants further study.

Finally, it is necessary to mention some important caveats to this work, and the construction of positive vacuum energy solutions in string theory in general.  First, recent work \cite{Moritz:2017xto} has shown that the antibrane uplift in the KKLT \cite{Kachru:2003aw} construction of de Sitter vacua with a single K\"ahler modulus is not sufficient to achieve positive vacuum energy.  This proof does not directly apply to time dependent backgrounds, or to the LVS stabilization mechanism we use, and therefore it presents no immediate obstacle to our model.  However, by showing that, in the case of a single K\"ahler modulus, the antibrane uplift does not simply add to the K\"ahler moduli potential, \cite{Moritz:2017xto} invites serious reservations as to the accuracy of all de Sitter constructions using antibranes. A better understanding of the positive energy added by antibranes in time dependent backgrounds with multiple K\"ahler moduli would certainly benefit the string cosmology community and be extremely relevant to this model.  Additionally, the recent work \cite{Sethi:2017phn} argues that the use of non-perturbative effects in the superpotential -- needed to stabilize K\"ahler moduli -- is not trustworthy in the presence of supersymmetry breaking flux.  This analysis casts suspicion on all known methods of K\"ahler moduli stabilization using non-perturbative effects.  While there are no no-go's which prevent stabilization, further study of non-perturbative effects in time dependent backgrounds is clearly an important topic for future research.

The plan for the remainder of this paper is as follows: stabilize, inflate, and backreact.  In \secref{modstab} we specify the set-up in which we will stabilize all moduli using both fluxes at tree-level, and also the leading $\alpha'$ corrections in the LVS potential.   In \secref{inflatingsector} we explain the inflationary mechanism and derive the inflationary potential. We also introduce the first effect of backreaction, taking into account the presence of antibranes in reducing the depth of the KS throat. In \secref{realizinginflation} we collect the long list of constraints on parameter space that must be respected in order for our approximations to be valid. Then, we present explicit realizations which respect these constraints and discuss their properties and observables. In \secref{Bkrxn} we discuss backreaction effects at length, explicitly computing effects due to the evolution of the K\"ahler moduli, as well as estimating the results of supersymmetry breaking on the asymptotic KS geometry.  Lastly, we discuss possible avenues for future research in \secref{Conclusions}.

\section{Stabilizing moduli}\label{modstab}
Before the inflationary dynamics of a four-dimensional low-energy effective theory descending from a string compactification can be considered, one must ensure the stabilization of the geometric moduli of the compact manifold. These moduli must receive masses above the four-dimensional Hubble scale so that they can be safely integrated out of the four-dimensional EFT. In this section we discuss the potentials which result in stable minima for the complex structure and K\"ahler moduli.  A discussion of the backreaction of the inflationary dynamics on these moduli will be presented in \secref{Bkrxn}.
\subsection{Complex structure moduli }
To achieve a hierarchy of scales between the compact manifold and a four-dimensional cosmology, we use the  warped orientifold compactifications of Giddings, Kachru and Polchinski \cite{Giddings:2001yu}.  In \cite{Giddings:2001yu} the authors show that all complex structure moduli can be stabilized by flux via the Gukov-Vafa-Witten tree-level superpotential \cite{Gukov:1999ya}:
\be
W_0 = \int_X G_3 \w \Omega~,
\ee
where $G_3 = F_3 - \tau H_3$, $\tau$ is the axio-dilaton and $\Omega$ is the holomorphic $(3,0)$-form of the internal manifold $X$.  Furthermore, the axio-dilaton, $\tau = C_0 +i \e^{-\phi}$, can be fixed by three-form flux. We will take $C_0=0$, and use a constant dilaton.  The values of $W_0$ and $g_s$ are fixed by global flux numbers and we will take them to be free parameters.  

The masses of the complex structure moduli are estimated by \cite{Kachru:2003aw} $m_{cs} \sim \alpha'/R^3$, where $R^6\sim \V$ is a typical length scale of the compact geometry (a precise definition of $\V$ is given in \eqref{metric}). We will require this mass to be larger than the Hubble constant during inflation: $m_{cs} \gg H$ such that the complex structure moduli remain stabilized and non-dynamical.  However, at tree-level, the compactifications of \cite{Giddings:2001yu} are `no-scale' models, meaning that the K\"ahler moduli are flat directions in moduli space.  In the next sub-section we will add non-perturbative and $\alpha'$ corrections in order to stabilize the overall volume. 

\subsection{K\"ahler moduli }\label{sec-LVS}
In the previous string embedding of unwinding inflation \cite{Gautason:2016cyp}, the KKLT model \cite{Kachru:2003aw} was employed to stabilize a single K\"ahler modulus corresponding to a four-cycle volume.  That treatment fell short in that the overall volume modulus, which determines the four-dimensional Planck mass, was not explicitly tied to the single stabilized four-cycle and was treated as an independent parameter.  Here, we will employ the minimal LVS scenario \cite{Balasubramanian:2005zx,Conlon:2005ki} with two K\"ahler moduli and the leading $\alpha'$ correction to the K\"ahler potential \cite{Becker:2002nn}.  This slight extension allows us to repair the leaky patches in the previous embedding. 

We start by specifying a string frame metric in which the stabilized volume modulus can be identified with a constant shift of the warp factor\cite{Giddings:2005ff,Frey:2008xw,Aparicio:2015psl}:
\begin{align}
\d s^2&=\mathcal{V}^{1/3}e^{2A(y)}ds_4^2 +e^{-2A(y)} ds_{{\rm CY}_0}^2~, \label{metric}\\
e^{-4A(y)}&=h(y)+\mathcal{V}^{2/3}~, \nonumber \\
ds_4^2 &= {g_4}_{\mu \nu}\d x^\mu \d x^\nu~, \nonumber \\
ds_{{\rm CY}_0}^2 &= \V^{-1/3} ds_{\rm CY}^2 = {g_{{\rm CY}_0}}_{mn}dy^mdy^n~. \label{unitCY}
\end{align}
In this ansatz, $g_4$ is the metric on four-dimensional spacetime, $ds_{{\rm CY}_0}$ is the line element on the compact Calabi-Yau rescaled so that it has unit volume\footnote{It is important to note a subtlety regarding \eqref{unitCY}: the rescaling using the volume modulus \emph{does not} rescale the radius of the $S^3$ at the tip of the KS throat which is a blow-up cycle whose size is fixed by fluxes \cite{Frey:2003dm}.}, $h(y)$ describes the warping, and $\V$ is the dimensionless volume modulus related to physical volumes via $\vol_6 \sim \V l_s^6 = \V (2\pi)^6 \alpha'^3$.  In what follows we assume that the compact manifold consists of a weakly warped bulk ($h_{bulk}\ll \V^{2/3}$) and a small warped throat region ($h_{KS}\gg \V^{2/3}$).  Using this ansatz to reduce to a four-dimensional Einstein frame we find that the Planck mass is given by:
\begin{equation} \label{MpofV}
\frac{M_{pl}^2}{2}=\frac{1}{2\kappa^2g_s^2}\left(\int\d^{6}y\sqrt{g_{\rm CY_0}}e^{-4A(y)}\mathcal{V}^{1/3}\right) \approx \frac{(2\pi)^3\alpha'^3\V}{2 \kappa^2 g_s^2}=\frac{\V}{2\pi \alpha' g_s^2}~,
\end{equation}
where the approximation is good in the limit where warping can be neglected over the majority of the compact manifold.  

We will assume a `strong Swiss-cheese' structure for the Calabi-Yau three-fold with only two K\"ahler moduli: 
\be \label{VofKaehler}
\mathcal{V} = \tau_{big}^{3/2} - \tau^{3/2} \approx \tau_{big}^{3/2}~.
\ee
Generally one should allow for a linear combination of moduli $\mathcal{V}= \alpha(\tau_{big}^{3/2} - \gamma \tau^{3/2})$. For simplicity, we set $\alpha,\gamma= 1$. We view this as conservative since including $\alpha$ and $\gamma$ as tuneable parameters only increases the flexibility of the model. In lieu of an explicitly constructed Calabi-Yau which accommodates our model\footnote{While there are explicit examples that match the structure \eqref{VofKaehler}, we additionally require a KS throat region, and large flux numbers that imply an F-theory uplift to a Calabi-Yau four-fold with extremely large Euler number.} this structure is motivated by the fact that stabilization of several moduli has been explicitly carried out in Swiss cheese examples using LVS \cite{Collinucci:2008sq}.  Further, it is argued in \cite{Anguelova:2009ht} that in fibred Calabi-Yau,  one expects to find vacua at large volume where a large number of moduli can be stabilized by higher order corrections in $g_s$. Therefore, we use the structure \eqref{VofKaehler} as a proof of principle for this model and assume that the specialization to an explicit example which accommodates a warped throat region is possible.

The string frame LVS K\"ahler and superpotential are given by:
\be \label{kahlersuper}
K= -2\log\(\V + \frac{\xi}{2}\) + \log\frac{g_s}{2} +K_{cs}~, \qquad W = W_0+\sum A_i \e^{-a_i \tau_{i}/g_s}~.
\ee
The K\"ahler potential includes the leading $\alpha'$ correction given by the $\xi/2$ term.  Here, $\xi$ is a positive constant that depends on the Euler number of the Calabi-Yau three-fold and $K_{cs} = \log\left(-i \int \Omega \w \overline{\Omega}\right)$. In the following, $\xi$ will be treated as a free parameter and $K_{cs} = 0$.  In the superpotential, $A_i$ are complex structure-dependent constants, and $a_i=2\pi/N_i$ where $N_i=1$ if the non-perturbative effect on the particular four-cycle $\tau_i$ arises from a Euclidean D3 brane, and $N_i=N_{D7}$ if the non-perturbative effect is gaugino condensation on a stack of D7 branes wrapped on $\tau_i$.

Because of the assumed hierarchy, $\tau_{big}\gg \tau$, non-perturbative effects are only relevant on the smaller four-cycle.  Then, the contribution of the LVS potential to the four-dimensional supergravity action is:
\be\label{LVSpot}
S_{LVS} \supset-\frac{g_s^4 M_p^4\e^{K_{cs}}}{8\pi} \int d^4x \sqrt{-g_4} \left[\frac{8}{3}\frac{(aA)^2}{g_s^2}\frac{\sqrt{\tau}e^{-2a\tau/g_s}}{\mathcal{V}} - 4aAW_0\frac{\tau e^{-a\tau/g_s}}{g_s \mathcal{V}^2} + \frac{3\xi W_0^2}{4\mathcal{V}^3}\right],
\ee
where we choose to parameterize the moduli using the small four-cycle, $\tau$, and the overall volume modulus, $\V\approx\tau_{big}^{3/2}$.
The pre-factor to the term in square brackets is necessary in order for the usual F-term potential arising from \eqref{kahlersuper} to match the result of dimensional reduction \cite{Anguelova:2009ht}. 

The LVS potential has an $AdS$ minimum for\footnote{The approximation here is due to the fact that $\frac{a\tau}{g_s} \gg1$, as needed in order to ignore higher order instanton effects.}:
\begin{align}
\langle \tau \rangle & \approx \left(\frac{\xi}{2 }\right)^{\frac{2}{3}}~, \\
\langle \mathcal{V}\rangle &\approx\frac{3g_s}{4 a A}W_0 \sqrt{\langle \tau \rangle} \e^{a \langle \tau \rangle/g_s}~.
\end{align}
Here, the moduli are stabilized using only the potential in square brackets in \eqref{LVSpot}, and only after a stable minimum is found should the overall potential be scaled by the factor $g_s^4 M_p^4/(8\pi)$. 

\begin{figure}
\begin{center}
\includegraphics[width=\textwidth]{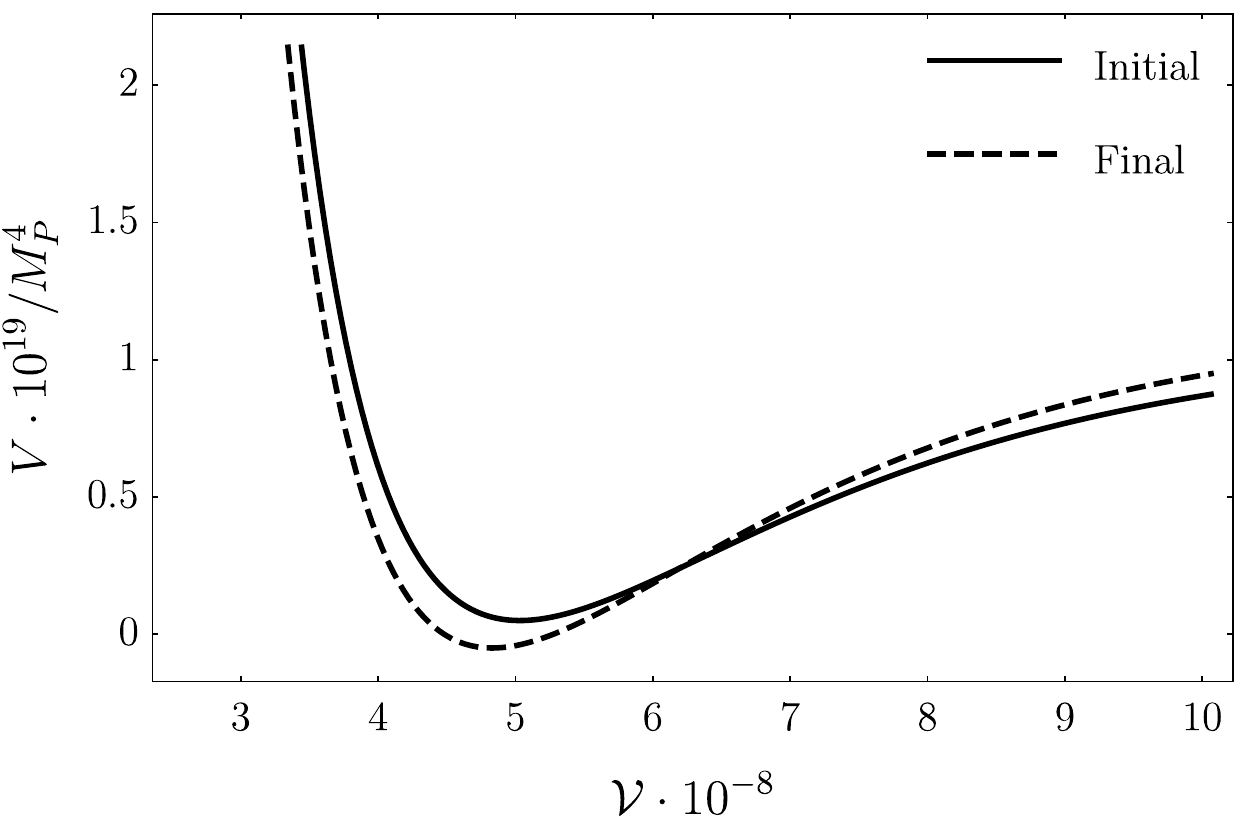}
\caption{The uplifted moduli potential \eqref{fullKaehlerPot} as a function of $\cal{V}$ at the beginning of the flux cascade and at the end. Plot made using parameter set 1 given in Table \ref{table-param}.}\label{fig-evolPotential}
\end{center}
\end{figure}

We will now lift this $AdS$ minimum by adding the  effect of the $p$ anti-D3 branes. The antibranes contribute to the K\"ahler moduli potential via a positive `uplift' that scales like $\V^{-4/3}$ \cite{Kachru:2003sx}.  The uplift potential is calculated in  the next section; the result is $V_{up}={2 p /( g_s h_0 \V^{4/3})}$.  Then, the expectation values of the K\"ahler moduli in the presence of $p$ antibranes is given by minimizing the following potential:
\be \label{fullKaehlerPot}
V(\tau, \V) = \frac{2 p}{g_s h_0 \V^{4/3}} + \frac{8}{3}\frac{(aA)^2}{g_s^2}\frac{\sqrt{\tau}e^{-2a\tau/g_s}}{\mathcal{V}} - 4aAW_0\frac{\tau e^{-a\tau/g_s}}{g_s \mathcal{V}^2} + \frac{3\xi W_0^2}{4\mathcal{V}^3}~, 
\ee
where $h_0$ represents the warp factor at the location of the antibranes. If $V_{up}$ is too large, one risks destabilizing the K\"ahler moduli. The beginning of the cascade, before any of the $p$ anti-D3 branes have annihilated against flux, is the most dangerous point for the stability of $\V$.  Then, the modulus $\V$ becomes more stable throughout the cascade as can be seen from Figure \ref{fig-evolPotential}.

In order to treat this model as single field inflation, we need the masses of the K\"ahler moduli to be greater than the Hubble parameter, and the expectation values not to evolve dramatically during inflation.  Using the results of \cite{Conlon:2007gk,Anguelova:2009ht}, the canonically normalized mass eigenstates corresponding to the two K\"ahler moduli of a strong Swiss cheese model are to leading order in a large $\V$ expansion:
\begin{align}
m^2_{\tilde{\tau}} &\approx \frac{g_s^4 \e^{K_{cs}}}{8\pi} \frac{18a^2g_s^3W_0^2\xi\sqrt{2\tau}}{\V^2}M_p^2~, \label{kahlermass1}\\
m^2_{\tilde{\tau_b}} &\approx \frac{g_s^4 \e^{K_{cs}}}{8\pi} \frac{729 g_s^6 \xi W_0^2}{8\sqrt{2} a \tau \V^3}M_p^2 \label{kahlermass2}~,
\end{align} 
where by $m_{\tilde{\tau}}$($m_{\tilde{\tau_b}}$) we indicate the mass eigenstate that is nearly aligned with the small(large) K\"ahler modulus.  These masses are reliable for order of magnitude estimates, however are subject to correction in two respects.  First, they are computed in the AdS minimum; uplifting the potential is expected to reduce $m_{\tilde{\tau_b}}$ by an order one factor and leave $m_{\tilde{\tau}}$ nearly invariant \cite{Conlon:2007gk}.  Second, the numerical factors are sensitive to the factors $\alpha,\gamma$ which we discuss under \eqref{VofKaehler}.  While we set $\alpha,\gamma=1$, the moduli masses could change dramatically if $\alpha,\gamma \not\approx 1$.  In addition to $m_{\tilde{\tau}} , \,m_{\tilde{\tau_b}} \gtrsim H$, if the moduli's expectation values evolve significantly over the inflationary epoch, backreaction effects will be strong.  We check moduli masses in \secref{Examples} and compute backreaction effects due to the evolution of the moduli vevs in \secref{kahlerbkrxn}.

\section{The inflating sector} \label{inflatingsector}
\subsection{The unwinding mechanism} \label{explainmechanism}
Having stabilized all moduli, we move on to the embedding of the unwinding inflation mechanism into warped throat geometries.  We refer the reader to \cite{D'Amico:2012ji} for more details on the general mechanism, and \cite{Gautason:2016cyp} for a more in-depth description of the specific realization used here.  The recipe for inflation calls for $p$  anti-D3 branes in a warped throat region of the compact geometry (we will study the scenario both in the KS throat and its s-dual (SDKS)).  Due to the presence of $F_5$ flux in the throat geometry, the anti-D3 branes are forced to the point of highest warping - the tip of the throat.  Once at the tip of the throat, the antibranes polarize via the Myers effect \cite{Myers:1999ps} into a 5-brane which wraps an $S^2$ localized in the polar direction, $\psi$, of the $S^3$ at the tip of the deformed conifold.  This 5-brane can be thought of as a bubble with decreased three-form flux (under which it is magnetically charged) in the interior.  Through the process of brane-flux annihilation \cite{Kachru:2002gs}, the reduction of three-form flux is accompanied by the reduction of antibrane charge.  

We will be interested in the case where this brane-flux annihilation has no fixed point, but rather results in an unstable potential for the 5-brane.  Then, the 5-brane bubble will expand in the $\psi$-direction and move across the compact space, removing three-form flux and decreasing $p$ as it goes.  This repeated brane-flux annihilation is an instance of a flux cascade \cite{Kleban:2011cs} which results in a slow roll potential. As in the original unwinding model, the canonically normalized inflaton is proportional to the distance in the $\psi$-direction that the 5-brane has moved.  A closely related model which uses a single instance of brane-flux annihilation (as opposed to flux cascade) to give rise to slow roll inflation can be found in \cite{DeWolfe:2004qx}.

In this section we reproduce the local embedding of the unwinding inflation mechanism in the KS geometry  and the SDKS geometry that was presented in \cite{Gautason:2016cyp}.  Furthermore, we extend the previous work to incorporate a unified treatment of global volume modulus stabilization with the inflationary dynamics. This results in an action describing the motion of the 5-brane on the $S^3$, \emph{i.e.} our inflaton action.  Ultimately, we prefer to embed the unwinding mechanism in SDKS because the stack of anti-D3 branes polarize into a D5, as opposed to an NS5 whose action is not known at weak string coupling.  However, for the sake of presentation, we begin with the familiar KS geometry, and then summarize the SDKS result afterwards.

\subsection{Unwinding in Klebanov-Strassler} \label{KSgeometry}
The literature on the KS geometry is both rich and vast, and we will refer the reader to previous works for a full appreciation of the solution.  Particularly, we have found \cite{Herzog:2001xk} to be a valuable resource and we will follow their conventions here. The KS solution describes a warped product of a four-dimensional spacetime with the deformed conifold, defined by $\sum_{i=1}^4 z_i^2 = \ep^2$, where $z_i$ are coordinates on $\mathds{C}_4$.  In the limit $\ep \to 0$ this is a singular cone over $T^{1,1}$; non-zero $\ep$ corresponds to blowing up the conical singularity on the $S^3$ of $T^{1,1}$.  

Striving towards a cosmologically viable model, we need to work in a compactified geometry.  Then, the KS throat is thought of as a localized region of a  compact Calabi-Yau \cite{Giddings:2001yu}.  In a compact manifold, the three-form flux in the KS throat is quantized  on both the blow-up $S^3$ and its dual, $\overline{S^3}$:
\be \label{quantizeflux}
\frac{1}{4\pi^2\alpha'}\int_{S^3} F_3 = M~, \qquad \frac{1}{4\pi^2\alpha'}\int_{\overline{S^3}} H_3 = K~.
\ee
The other relevant details of this solution are:
\be
\begin{split}
\tau &= \frac{i}{g_s}~, \qquad G_3 = F_3 - \tau H_3~, \qquad \star_6 G_3 = iG_3~, \\
C_4 &= \beta \vol_4~, \qquad h(y)^{-1} g_s^{-1} = \beta~.
\end{split}
\ee
The integrated Bianchi identity for the five-form flux, $F_5 = \d C_4$, gives rise to a tadpole condition for the three-form flux and D3 charges:
\be \label{tadpole}
 \frac{1}{(4\pi^2 \alpha')^2} \int H_3 \w F_3 + N_{D3} = \frac{\chi}{24}~,
 \ee
 where $N_{D3}$ counts the localized D3 charge and $\chi$ is the Euler number of the corresponding fourfold in F-theory \cite{Giddings:2001yu}.
 
Probe anti-D3 branes in this geometry will feel a force due to the presence of $F_5$ flux which drives them to the region of highest warping - the bottom of the KS throat.  Therefore, the flux cascade is localized and confined at the bottom of the throat where $h(y)\gg \V$ and the metric is of the form \cite{Herzog:2001xk}:
\begin{align}
&\d s_{tip}^2=\mathcal{V}^{1/3}h_0^{-1/2}{g_4}_{\mu \nu}\d x^\mu \d x^\nu+b_0^2g_s M \alpha'  ( \d \Omega_3^2 + \frac{1}{4\V^{1/3}} \d \rho^2 + \frac{1}{16\V^{1/3}} \rho^2\d \Omega_2^2)~,\label{metricKSTip} \\
&h(\rho=0)^{1/2}\equiv h_0^{1/2}=\left(\frac{3}{2}\right)^{1/3}b_0^2g_s M\alpha'\epsilon^{-4/3}~,
\end{align}
where $\rho$ is a dimensionless radial coordinate which is equal to zero at the tip of the throat and  $b_0^2\approx 0.933$.  Note that the  size of the blow-up $S^3$ at the tip of the throat does not scale under \eqref{unitCY} -- that is, the size of a blow-up cycle is fixed by fluxes, regardless of the size of the manifold it is attached to.

In a compact manifold, the warped hierarchy between the tip (IR) and top (UV) of the throat is controlled by the deformation parameter, $\ep$.  In the absence of probe branes, this parameter is given by \cite{Giddings:2001yu}:
\be \label{eps0}
\ep_{(0)}^{2/3} = r_{\rm UV} \e^{-2\pi K/(3g_s M)}~,
\ee
where the $(0)$ subscript is used to refer to $0^{th}$ order in the probe approximation.  We also introduce the radial coordinate, $r$, which is related to $\rho$ at large $\rho$ via $r^2=3\ep^{4/3}\exp(2\rho/3)/2^{5/3}$, and by $r_{\rm UV}$ we denote the location where the throat transitions  into the bulk geometry. 

In the large $r$ limit the KS metric is given by  \cite{Herzog:2001xk}:
\begin{align}
\d s_{\rm UV}^2 &= \mathcal{V}^{1/3}h_{\rm UV}^{-1/2}{g_4}_{\mu \nu}\d x^\mu \d x^\nu+\frac{h_{\rm UV}^{1/2}}{\V^{1/3}} ( \d r^2 + r^2 \d s_{T_{1,1}})~,  \label{metricUV} \\
h(r\to \infty) &\equiv h_{\rm UV}= \frac{L^4}{r^4}\(\log\frac{r}{\ep^{2/3}}-\frac{1-\log(1024/729)}{12}\)~, \qquad L^4 = \frac{81 (g_s M\alpha')^2}{8}~. \label{lgrwarp}
\end{align}
The numerical factors in $h_{\rm UV}$ are insignificant and will be dropped henceforth.
This UV expression matches the asymptotics of the singular Klebanov-Tseytlin (KT) solution \cite{Klebanov:2000nc}:
\begin{align}
\d s_{\rm KT}^2 &= h_{\rm KT}^{-1/2}{g_4}_{\mu \nu}\d x^\mu \d x^\nu+h_{\rm KT}^{1/2}  ( \d r^2 + r^2 \d s_{T_{1,1}})~, \\
h_{\rm KT}(r) &= \frac{L^4}{r^4} \log \frac{r}{r_s} \\
&= \frac{L^4}{r^4}\( \log \frac{r}{r_{\rm UV}} + \frac{2\pi N}{3g_s M^2} +\frac14 \)~,
\qquad r_s = r_{\rm UV}\, \e^{\frac{-2\pi N}{3g_s M^2}-\frac14}~,
\end{align}
which is a non-compact solution with  a singularity at $r=r_s$, containing both three form flux and $N$ D3-branes. To match the asymptotics of  the two solutions (neglecting insignificant numerical factors) one should associate the location of the singularity in KT with the scale of the deformation in KS, $r_s \to \ep^{2/3}$, and the number of branes in KT should be replaced by the total D3 charge in the compact KS geometry, $N=KM-p$.  This asymptotic charge matching takes into account the the presence of the probe antibranes as they lower the net D3 charge responsible for the depth of the throat.  The resulting deformation parameter is \cite{DeWolfe:2008zy}:
\be \label{epsKS}
\ep^{2/3} = r_{\rm UV} \e^{\frac{-2\pi}{3g_s M^2}( KM-p)}~,
\ee
which  reduces to \eqref{eps0} when the probe approximation is valid: $p\ll KM$.  Following \cite{Baumann:2007ah}, we define the place where the throat attaches to the bulk manifold, $r_{\rm UV}$, as the radial position where the warp factor \eqref{lgrwarp} is unity; this defines:
\be \label{ruv}
r_{\rm UV}^4 = \frac{27\pi g_s {\alpha'} ^2}{4}\( KM - p \)~.
\ee 

Taking into account the presence of the probe branes in the warp factor constitutes the leading order backreaction on the UV geometry and is crucial to the reliability of our inflationary model.  We will discuss this further, as well as additional backreaction effects, in \secref{Bkrxn}.

\paragraph{NS5 inflaton action}\mbox{}\\
With these details in hand, we can compute the action for the probe NS5-brane formed via the polarization of $p$ anti-D3 branes \cite{Kachru:2002gs}:
\begin{equation}
S_{NS5} = -\frac{\mu_5}{g_s^2} \int d^6 \xi [-\det(G_\parallel)\det(G_\perp +2\pi g_s  {\cal F}_2 )]^{1/2} -\mu_5 \int B_6~,
\end{equation}
with
\begin{equation}\label{formFields}
2\pi  {\cal F}_2=2\pi \sqrt{\alpha'} F_2-C_2~, \quad \int_{S^2}F_2=2\pi \sqrt{\alpha '}p~.
\end{equation}
The stack of $p$ anti-D3 branes polarize into a bubble that is extended in a four-dimensional FLRW cosmology and the polar angle on the $S^3$, while $G_\perp$ describes the $S^2$ on the $S^3$ which the bubble wraps.  For the details of the computations see \emph{e.g.} \cite{Kachru:2002gs,DeWolfe:2004qx,Gautason:2016cyp}.  The resulting action, taking into account the volume modulus for the compact geometry \eqref{metricKSTip}, is:
\be \label{NS5action}
\begin{split}
S_{NS5} &=-\frac{M\mathcal{V}^{2/3}h_0^{-1}}{8\pi^4 g_s \alpha'^2} \int\d^4xa^3(t)\left[V_2(\psi)\sqrt{1-\frac{b_0^2g_sM \alpha' h_0^{1/2}}{\mathcal{V}^{1/3}}\dot{\psi^2}}+U(\psi)\right]~, \\
&= -\frac{M_p^4 g_s^4}{8 \pi} {\cal C} \int\d^4xa^3(t)\left[V_2(\psi)\sqrt{1-Z^2\dot{\psi^2}}+U(\psi)\right]~,
\end{split}
\ee
where we use $\mu_3=(2\pi)^2\alpha' \mu_5=(2\pi)^{-3} \alpha'^{-2}$ and define:
\begin{equation}
\begin{split}
&V_2(\psi)=\sqrt{b_0^4\sin^4(\psi)+U^2(\psi)}~, \qquad U(\psi)= \frac{\pi p}{M} - \psi + \frac12\sin(2\psi)~, \\
& {\cal C}=\frac{M h_0^{-1}}{g_s\pi\V^{4/3}}~, \qquad Z^2=\frac{b_0^2g_sM \alpha' h_0^{1/2}}{\mathcal{V}^{1/3}}~.
\end{split}
\end{equation}
Here, we use \eqref{MpofV} to pull out an overall factor $M_p^4 g_s^4 / 8 \pi$ in order to identify the uplift term, $V_{up}=2{\cal C}U(\psi)$, in the K\"ahler potential \eqref{fullKaehlerPot}.

\begin{figure}
\begin{center}
\includegraphics[width=\textwidth]{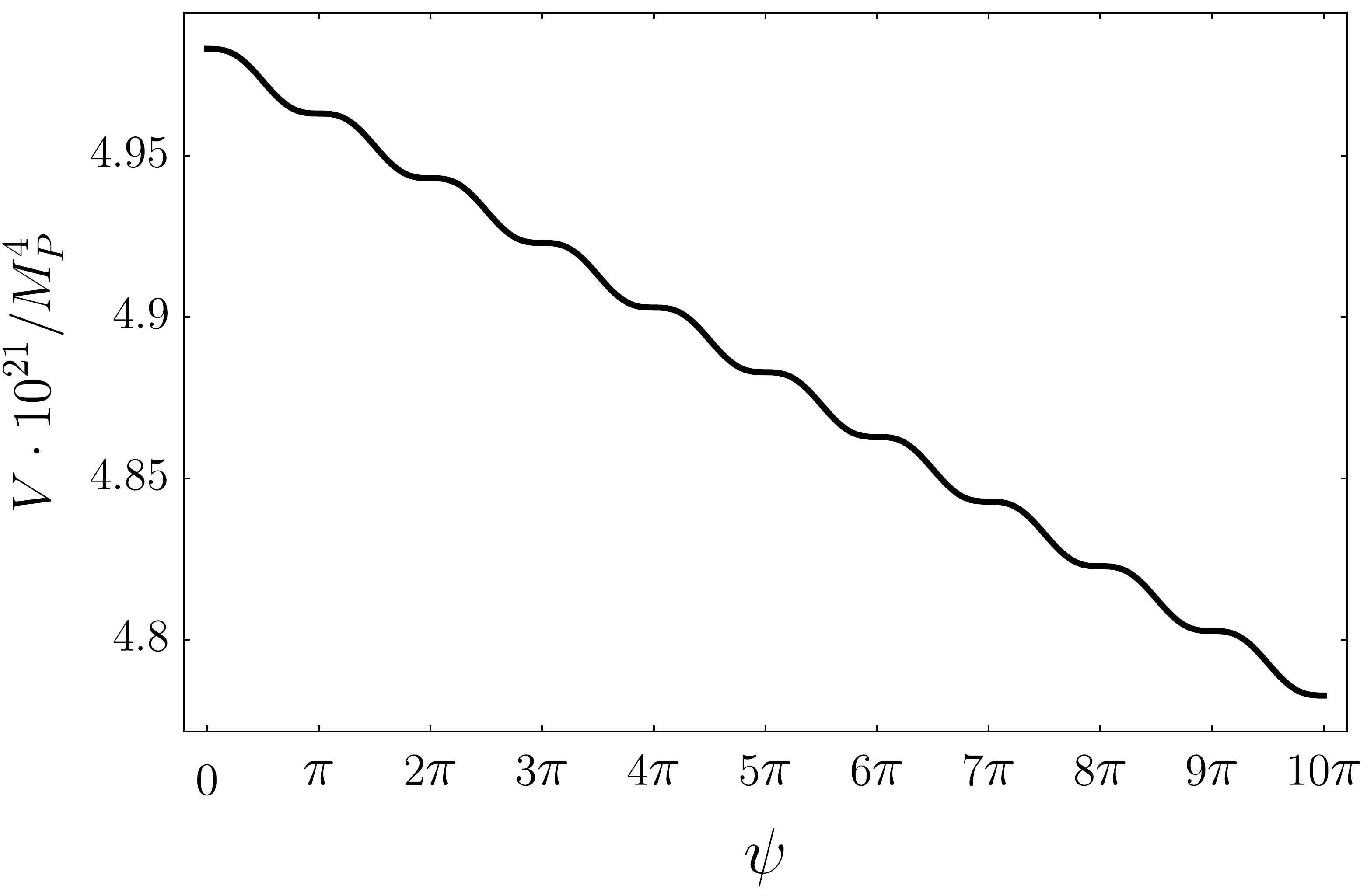}
\caption{The inflaton potential as a function of the 5-brane position on the $S^3$.}\label{fig-psi-potential}
\end{center}
\end{figure}

Expanding the DBI  kinetic term in \eqref{NS5action} in both small velocity and large\footnote{Large $p/M$ is the regime of parameter space where a flux cascade is possible.} $p/M$  we identify the canonically normalized inflaton, $\phi$:
\be \label{canfieldKS}
\phi = \frac{M_p^2g_s^2}{\sqrt{8\pi}}\, \frac{b_0 h_0^{-1/4} \sqrt{p M \alpha'}}{\V^{5/6}}\psi \equiv f(\V)\psi~.
\ee
This expression assumes that the volume modulus $\V$ remains constant and so is only valid in the regime of small backreaction.  Taking into account the backreaction of inflation will also alter the field range - we compute the effect of this backreaction in \secref{kahlerbkrxn}.
In terms of the canonically normalized field, we can write the full inflaton potential, including the LVS vacuum contribution \eqref{LVSpot}:
\be \label{canpotKS}
S =\int d^4x a^3(t) \( \frac12 \dot{\phi}^2 - \frac{M_p^4 g_s^4}{8\pi} \big( {\cal C}(V_2(\phi) +U(\phi) )+ \Lambda_{LVS}\big) \)~.
\ee
The potential, which is linear with oscillations is shown in \figref{fig-psi-potential}.

\subsection{Unwinding in the S-dual of Klebanov-Strassler} \label{unwindingSDKS}
Taking the S-dual of the KS geometry  switches the positions of the $F_3$ and $H_3$ flux, such that the stack of anti-D3 branes polarize into a D5 brane as opposed to an NS5 brane.  It should be noted that while S-duality is a strong-weak duality, taking $g_s\to g_s^{-1}$, we use it as a solution generating technique, and then study the new solution at weak coupling. Therefore the two cases we look at, KS vs SDKS, are not identical as they are both studied at weak coupling. Switching $F_3$ and $H_3$ takes  \eqref{quantizeflux} to:
\be
\frac{1}{4\pi^2\alpha'}\int_{\overline{S^3}} F_3 = M~, \qquad \frac{1}{4\pi^2\alpha'}\int_{S^3} H_3 = K~.
\ee
Using tildes to indicate quantities are given in the S-dual geometry, the form of the metric at the tip of the SDKS throat is \cite{Herzog:2002ih}:
\begin{align}
&\d s^2=\mathcal{V}^{1/3}\tilde{h}_0^{-1/2}{g_4}_{\mu \nu}\d x^\mu \d x^\nu+b_0^2 K\alpha' ( \d \Omega_3^2 + \frac14 \d \rho^2 + \frac{1}{16} \rho^2\d \Omega_2^2)~, \label{metricSDKSTip}\\
&\tilde{h}_0^{1/2}=\left(\frac{3}{2}\right)^{1/3}b_0^2g_s^{-2} K \alpha'\tilde{\epsilon}^{-4/3}~.
\end{align}
Following steps identical to those of \secref{KSgeometry} but now matching to the S-dual of the Klebanov-Tseytlin geometry, the deformation parameter becomes:
\be \label{epsSDKD}
\tilde{\ep}^{2/3} = \tilde{r}_{\rm UV}\, \e^{\frac{-2\pi g_s}{3K^2}(KM-p)}~,
\ee
and the warp factor at large $r$ is equal to unity at:
\be \label{SDruv}
\tilde{r}_{\rm UV}^4 = \frac{27\pi g_s{\alpha'}^2}{4}\(KM-p\)~.
\ee

\paragraph{D5 inflaton action}\mbox{} \newline
The action for a probe D5-brane is:
\begin{equation}
S_{D5} = \frac{-\mu_5}{g_s} \int d^6 \xi [-\det(G_\parallel)\det(G_\perp + 2\pi {\cal F}_2 )]^{1/2} -\mu_5 \int 2\pi{\cal F}_2\w C_4~,
\end{equation}
with 
\begin{equation}
2\pi  {\cal F}_2=2\pi \sqrt{\alpha'} F_2-B_2~, \quad \int_{S^2}F_2=2\pi \sqrt{\alpha '}p~.
\end{equation}
The D5 action in the S-dual of the KS geometry was calculated in \cite{Gautason:2016cyp}, following closely the original computation for an NS5-brane in KS \cite{Kachru:2002gs}.  Again, accounting for the correct treatment of the volume modulus \eqref{metricSDKSTip}, we find:
\be \label{D5action}
\begin{split}
S_{D5}&=-\frac{K\mathcal{V}^{2/3}\tilde{h}_0^{-1}}{8\pi^4 g_s\alpha'^2}\int\d^4x a^3(t)\left[\tilde{V}_2(\psi)\sqrt{1-\frac{b_0^2K\alpha'\tilde{h}_0^{1/2}}{\mathcal{V}^{1/3}}\dot{\psi^2}}+\tilde{U}(\psi)\right]~, \\
&= -\frac{M_p^4 g_s^4}{8 \pi} \tilde{{\cal C}} \int\d^4xa^3(t)\left[\tilde{V}_2(\psi)\sqrt{1-\tilde{Z}^2\dot{\psi^2}}+\tilde{U}(\psi)\right]~,
\end{split}
\ee
with analogous definitions:
\begin{equation}
\begin{split}
&\tilde{V}_2(\psi)=\sqrt{b_0^4\sin^4(\psi)+\tilde{U}^2(\psi)}~, \qquad \tilde{U}(\psi)= \frac{\pi p}{K} - \psi + \frac12\sin(2\psi)~, \\
& \tilde{{\cal C}}=\frac{K \tilde{h}_0^{-1}}{g_s\pi\V^{4/3}}~, \qquad \tilde{Z}^2=\frac{b_0^2 K \alpha' \tilde{h}_0^{1/2}}{\mathcal{V}^{1/3}}~.
\end{split}
\end{equation}

Expanding the action for the canonically normalized field, we find the SDKS analogues of \eqref{canfieldKS} and \eqref{canpotKS}:
\be \label{canfieldSDKS}
\phi = \frac{M_p^2g_s^2}{\sqrt{8\pi}}\, \frac{b_0 \tilde{h}_0^{-1/4} \sqrt{p K \alpha'}}{\sqrt{g_s} \V^{5/6}}\psi \equiv \tilde{f}(\V)\psi~,
\ee
and
\be \label{canpotSDKS}
S =\int d^4x a^3(t) \( \frac12 \dot{\phi}^2 - \frac{M_p^4 g_s^4}{8\pi} \big( \tilde{{\cal C}}(\tilde{V}_2(\phi) +\tilde{U}(\phi) )+ \Lambda_{LVS}\big) \)~.
\ee

\section{Realizing inflation} \label{realizinginflation}
In this section we will present several realizations of inflationary epochs. Before moving to specific scenarios, we will accumulate the various constraints on parameters that must be satisfied for the consistency of the description we have presented thus far.

\subsection{Constraints on parameter space} \label{constraints}
\begin{itemize}
\item \textbf{Validity of the string loop and $\alpha'$ expansion:} The assumptions of supergravity require: 
\begin{align}
g_s &\ll 1~, \nonumber \\
M,K &\gg 1~,  \nonumber \\
\V &\gg \xi~. \nonumber
\end{align}
\item \textbf{Validity of the probe approximation:} To ensure the antibrane charge is not so large that it backreacts significantly on the throat geometry we require:
\be
p \ll MK~. 
\ee
We also need to ensure that the polarized 5-brane does not significantly alter the $S^3$ at the tip of the throat. This will be satisfied as long as the radius of backreaction of the antibranes is small compared to the radius of the $S^3$:
\begin{align}
g_s p &\ll (g_s M)^2 \qquad \textrm{KS}~, \\
g_s p &\ll K^2 \qquad \textrm{SDKS}. 
\end{align}
\item \textbf{Validity of the non-perturbative expansion in the superpotential:} In order to safely truncate higher order non-perturbative effects we require:
\be 
\e^{-a\tau /g_s}\ll1~.
\ee
\item \textbf{Validity of the simplifying assumptions on the geometry:} Beginning with \eqref{metric}, we assume a small, strongly warped throat and a large, unwarped bulk.  The assumption of strong warping implies:
\be
h_0 \gg \V^{2/3}~.
\ee
Then, in order to safely neglect the contribution of the warped throat to the total warped volume in \eqref{MpofV} we require:
\begin{align}
\V&\gg \V_{throat} =\vol(T^{1,1})\V^{-2/3}\int_0^{r_{UV}} dr\, r^5 h_{UV}(r)~,  \\
\V&\gg \frac{3^{3/2}\pi^{7/2}  \sqrt{g_s (K M-p)}}{8 \V^{2/3} } \times
 \begin{cases}
g_s\big(8\pi (K M -p) -7g_s M^2\big)  & \textrm{KS,} \nonumber\\
\big(8\pi (K M -p) -7K^2\big) & \textrm{SDKS.} \nonumber
\end{cases}
\end{align}
Although for simplicity of the presentation we estimate the warped volume of the throat using the large $r$ form of the metric, it is clear from the expression that the majority of the volume comes from large $r$. The difference between this estimate and a numerical integration using the full KS warp factor has been checked and is negligible.
\item \textbf{Validity of single field inflation (cosmological hierarchy):} As mentioned in \secref{modstab} in order for the moduli to remain stabilized and non-dynamical during inflation, it is necessary that they be heavier than the Hubble scale:
\begin{align}
&\frac{1}{\V} > H^2~, \\
&m^2_{\tilde{\tau}} > H^2~, \\
&m^2_{\tilde{\tau_b}} > H^2~,
\end{align}
where estimates for the K\"ahler moduli masses are given in \eqref{kahlermass1} and \eqref{kahlermass2}.
\end{itemize}

\subsection{Examples} \label{Examples}
We present here a few examples of specific realizations of inflationary scenarios.  We choose to focus on the unwinding mechanism in SDKS where the antibranes polarize into a D5 as opposed to an NS5, whose action is less well-understood at weak coupling.  To avoid mixing KS and SDKS quantities in this section, we relegate an example that works in KS to Appendix \ref{app-KS}.

We find that in order to have successful inflation we are driven to the regime of large flux numbers, $M$ and $K$, and also, as is usually the case in LVS, large $W_0$. These large values of $M,K$ enter into \eqref{tadpole} to imply Euler characteristics for the associated Calabi-Yau four-fold that are larger than 1,820,448, which is the largest known \cite{Klemm:1996ts}.  Thus, we either need to rely on the existence of currently unknown manifolds with large Euler characteristic, or we need to assume that negative contributions to the left hand side of \eqref{tadpole} exist outside of the throat geometry.  Neither of these are entirely satisfactory, however both are technically possible.

\begin{figure}[H]
\begin{center}
\includegraphics[width=0.46\textwidth]{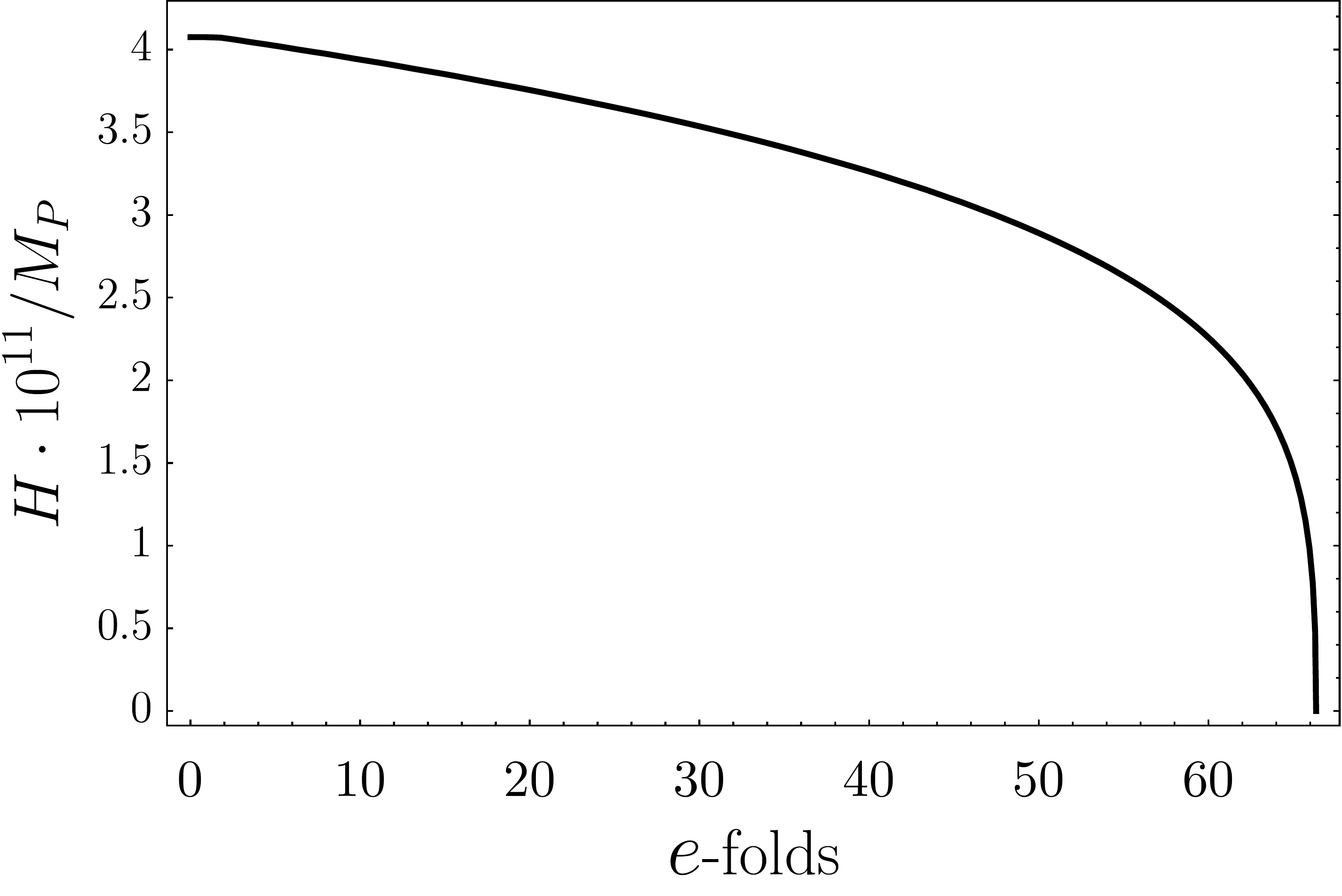}
\includegraphics[width=0.46\textwidth]{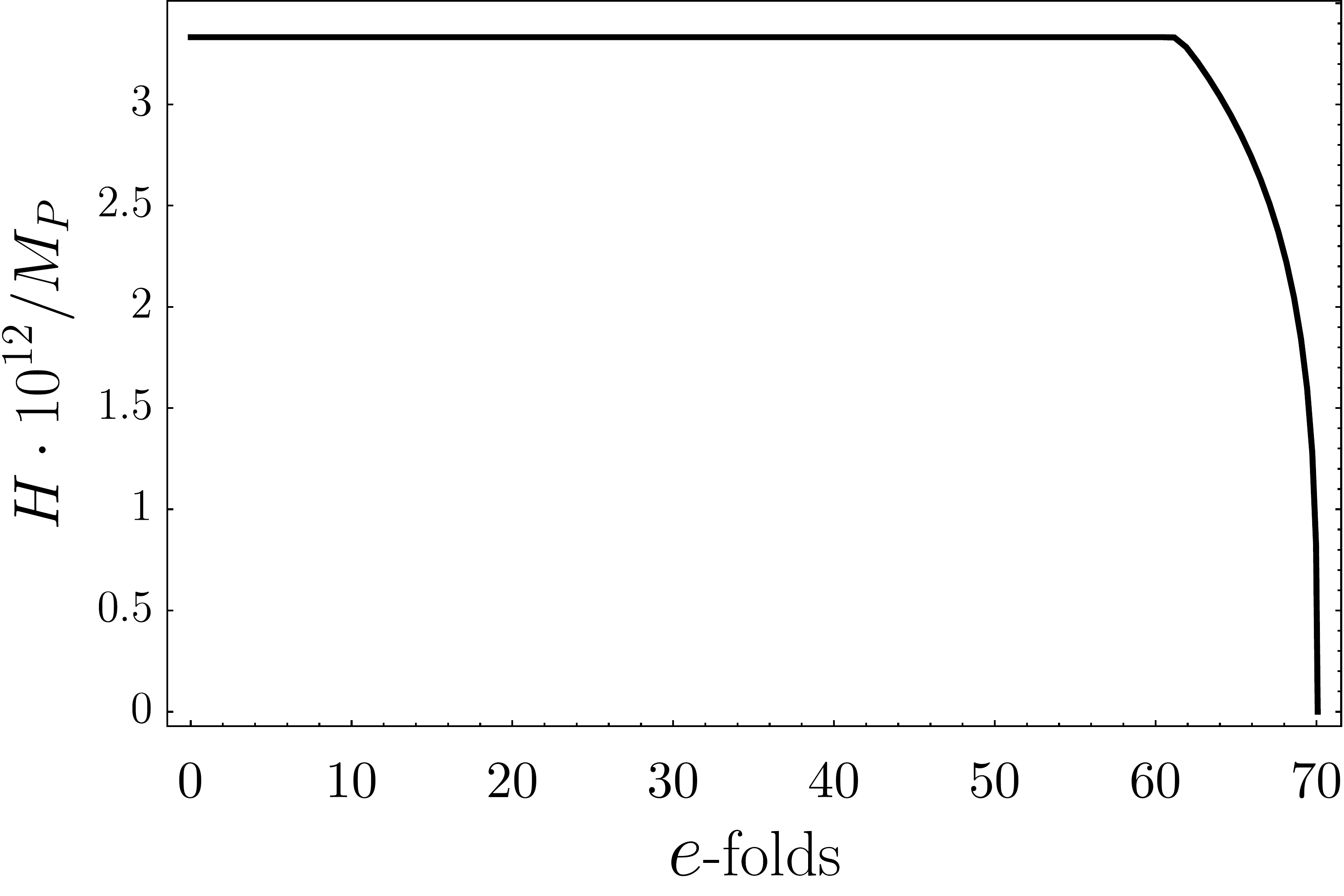}
\caption{Evolution of the Hubble parameter for set 1 (left) and set 2 (right).}\label{fig-Hubble}
\end{center}
\end{figure}

In Table \ref{table-param} we present two sets of parameters which give rise to qualitatively different inflationary scenarios.  The first, set 1, is very similar to the findings in \cite{Gautason:2016cyp}.  Although of less phenomenological interest due to a very small power spectral amplitude, it has a very large field range.  Meanwhile, set 2 represents a new two-phase version of unwinding inflation, which in addition to a super-Planckian field excursion also reproduces the correct amplitude of the scalar power spectrum.  Here, most of the efolds of inflation come from a single inflection point region of the potential near $\psi=0$\footnote{Caveat: because the inflaton lingers for many efolds near one pole, this scenario could potentially receive large corrections beyond the probe approximation due to open questions regarding the IR backreaction of antibranes, see \secref{Bkrxn}}. Then, the unwinding mechanism, still within the slow roll regime, exits inflation giving rise to the large field excursion, but only a few efolds.  The different evolution of the inflaton for these two sets can be seen in the evolution of the Hubble parameter in Figure \ref{fig-Hubble}.  Their power spectra and field ranges are collected in Table \ref{table-field+pspec}.  

In both scenarios the initial conditions for inflation place the polarized 5 brane near one pole with small velocity.  We require that the canonically normalized field displacement at the initial condition is larger than $H$, as one would naturally expect thermal fluctuations to be of this order; for the results listed in Table \ref{table-field+pspec} we take $\phi(0)\sim10^7H$ and $\phi(0)\sim10^5H$ for set 1 and 2 respectively. The unwinding scenario resulting from set 1 is insensitive to initial conditions, whereas in set 2 the number of efolds will increase if the initial condition is taken closer to the inflection point at $\psi=0$.

It is extremely non-trivial that there remains a viable region in the parameter space bounded by the highly non-linear constraints listed in \secref{constraints}  In Table \ref{table-constraints} we present the degree to which sets 1 and 2 satisfy these constraints.  Of course, seeing that a constraint is satisfied does not necessarily mean that it is satisfied to a strong enough degree to avoid catastrophic backreaction.  Thus, in \secref{kahlerbkrxn} we compute backreaction effects and summarize their magnitudes for these parameter sets in Table \ref{table-bkrxn}.

\begin{table}[H]
\begin{center}
\begin{tabular}{|c|c|c|c|c|c|c|c|c|c|c|}
\hline 
• & $a$ & $\xi$ & $A$ & $W_0$ & $g_s$ & $M$ & $K$ & $p$ & $\tau$ & $\cal{V}$  \\ 
\hline 
Set 1 & $\pi$ & 1.04 & 1.1 & $10^6$ & .23 & 38\,350 & 2\,700 & 10\,354\,500 & .694 & $5.03\cdot 10^8$  \\ 
\hline 
Set 2 & $\pi$ & 1.31 & 1 & $10^6$ & .23 & 38\,460 & 2\,300 & 12\,636\,857 & .801 & 2.59$\cdot 10^9$  \\ 
\hline 
\end{tabular}
\end{center}
\caption{Two sets of parameters which give rise to two different inflationary regimes.}\label{table-param}
\end{table}

\begin{table}[H]
\begin{center}
\begin{tabular}{|c|c|c|c|c|c|}
\hline 
• & $e$-folds & $\Delta\phi/M_P$ & $\mathcal{P}_\zeta$ & $n_s$ & $r$ \\ 
\hline 
Set 1 & 66 & 10.9 & 3.27$\cdot10^{-21}$ & .975 & .101 \\ 
\hline 
Set 2 & 69 & 3.62 & 1.26$\cdot 10^{-9}$ & 1.00 & 2.5$\cdot 10^{-15}$ \\ 
\hline 
\end{tabular}
\end{center}
\caption{Field range and power spectrum for both parameter sets. }\label{table-field+pspec}
\end{table}

\begin{table}[H]
\begin{center}
\begin{tabular}{|c|c|c|c|c|c|c|c|c|}
\hline 
• & $\frac{p}{KM}$ & $\frac{g_sp}{K^2}$ & $e^{-a\tau/g_s}$ & $\frac{\mathcal{V}^{2/3}}{h_0}$ & $\frac{\mathcal{V}_{throat}}{\mathcal{V}}$ & $\mathcal{V}H^2$ & $\frac{H^2}{m_{\overline{\tau}}^2}$ & $\frac{H^2}{m_{\overline{\tau}_b}^2}$ \\ 
\hline 
Set 1 & .100 & .327 & 7.59$\cdot10^{-5}$ & 1.94$\cdot10^{-6}$ & .253 & 2.53$\cdot10^{-3}$ & 1.19$\cdot10^{-6}$ & .647  \\ 
\hline 
Set 2 & .143 & .549 & 1.76$\cdot10^{-5}$ & 3.28$\cdot10^{-7}$ & 1.22$\cdot10^{-2}$ & 4.49$\cdot10^{-4}$ & 4.32$\cdot10^{-7}$ & .592  \\ 
\hline 
\end{tabular}
\end{center}
\caption{All the constraints mentioned in \secref{constraints} are satisfied by both parameter sets.  }\label{table-constraints}
\end{table}

\section{Backreaction} \label{Bkrxn}
The flux cascade changes the number of three-form flux as well as the number of antibranes, both of which contribute to the stabilization of the geometric moduli. Thus, through the dependence $N_{\overline{D3}}(\psi)$ and $K(\psi)$ (or $M(\psi)$ in SDKS), the moduli vevs depend on the inflaton.  This dependence will in turn lead to new forces on the inflaton which could spoil the flatness of the inflaton potential.  In this section we will discuss the backreaction on the geometry due to the presence and depletion of antibranes.  For simplicity, we will only consider the backreaction of the s-wave, or $0^{th}$ harmonic, an approximation that becomes better at large radius. First, we will take into account the backreaction due the presence of antibrane charge, but without taking into account supersymmetry breaking.  This can be though of as treating $p$ anti-D3 branes as $-p$ D3 branes (see \emph{e.g.}\,\,\cite{Flauger:2009ab}). We will restrict ourselves to an asymptotic charge matching along the lines of \cite{DeWolfe:2008zy}, without attempting to take sub-leading corrections to the UV expansion of the metric into account. Then, we will compute the contribution to the slow roll parameters due to the backreaction of the inflaton on the K\"ahler moduli vevs.  Finally, we estimate the additional effects of the supersymmetry breaking on the geometry as well as discuss open questions in the literature regarding IR backreaction effects.       

\subsection{Backreaction of antibranes on net D3 charge}
The first backreaction effect we wish to consider is that of the antibranes on the warping of the throat geometry. As explained in \secref{KSgeometry} and \secref{unwindingSDKS}, the matching of the asymptotic charges results in \eqref{epsKS} and \eqref{epsSDKD}, from which one can see that the presence of the antibranes reduces the net D3 charge, which in turn reduces the `depth' or total warping of the throat as seen from the UV.  Because of the reduced warping, each antibrane contributes a stronger uplift in the K\"ahler potential, so that not as many anti-D3s can be added before destabilizing the volume modulus.  This limitation on the number of antibranes that we can add ultimately limits the amount of inflation that can be realized in a given throat geometry.  

However, this backreaction is also crucial in that now the total warping only depends on the net D3 charge in the throat - a quantity that is conserved.  Specifically, the warping, \eqref{epsKS}, \eqref{epsSDKD}, only depends on the combination $KM-p$ which is exactly constant, as required by the tadpole condition \eqref{tadpole}, throughout inflation\footnote{The total warp factor $h_0$ ($\tilde{h_0}$) additionally depend on $M$ ($K$), but this flux is constant throughout the cascade in KS (SDKS).}.  The fact that the $\psi$-dependence cancels out of the warp factor means that no new terms in the inflaton potential will arise due the presence of the warp factor in the action.  Furthermore, the results of \cite{Baumann:2006th} show that beyond explicit dependencies on the warp factor in the inflaton potential, the non-perturbative effects in the superpotential depend exponentially on the warped volume of the four-cycle supporting the wrapped D7 or Euclidean D3 branes. Backreaction due inflaton dependence in this warped volume was computed for D-brane  inflation in \cite{Baumann:2007ah} and for axion monodromy in \cite{Flauger:2009ab}. Because the warped volume of the four-cycle appears exponentially in the superpotential, these effects are is especially dangerous.  The fact that the leading UV behavior of the warp factor is independent of the inflaton protects this model from potentially serious backreaction effects.

While in this approximation the inflaton dependence exactly cancels, this only indicates that the leading contribution to the $\psi$-dependence of the metric will enter as a higher harmonic.  Not only are higher harmonics generally suppressed exponentially with distance, but  the magnitude of this backreaction is not expected to grow throughout inflation due to charge conservation.  Although the leading backreaction beyond the s-wave approximation is beyond the scope of this work, charge conservation implies that it will not be a cumulative effect and should not significantly alter our results.  Furthermore, unlike large field inflation models in more complicated geometries, the calculation of this backreaction using  harmonic functions in KS \cite{Krishnan:2008gx} is relatively straightforward.

\subsection{Backreaction on K\"ahler moduli} \label{kahlerbkrxn}

Despite the cancelation of inflaton dependence in the warp factor, there will still be backreaction due to the dependence on the number of antibranes -- as opposed to the total D3 charge -- in the K\"ahler moduli potential \eqref{fullKaehlerPot}.  We will now check the backreaction on the slow roll parameters:
\be
\varepsilon = \frac{M_p^2}{2} \(\frac{V'(\phi)}{V(\phi)} \)^2\qquad {\rm and} \qquad \eta=M_p^2 \frac{V''(\phi)}{V}~,
\ee 
due to the dependence $M_p(\V)$,  $\tilde{{\cal C}}(\V)$, $\tilde{f}(\V)$ and $\Lambda_{LVS}(\V)$ and the backreaction $\V(\phi)$ and $\tau(\phi)$ for the inflaton potential \eqref{canpotSDKS}.  For simplicity we will only worry about the average contribution to the slow roll parameters.  Specifically, because the potential is linear with oscillations (see \figref{fig-psi-potential}), $\eta$ oscillates around zero with a large amplitude.  This oscillating part may give rise to resonant features in the power spectrum or non-Gaussianity, but is not important to the question of backreaction because it averages to zero over several efolds.

\begin{figure}[H]
\begin{center}
\includegraphics[width=0.45\textwidth]{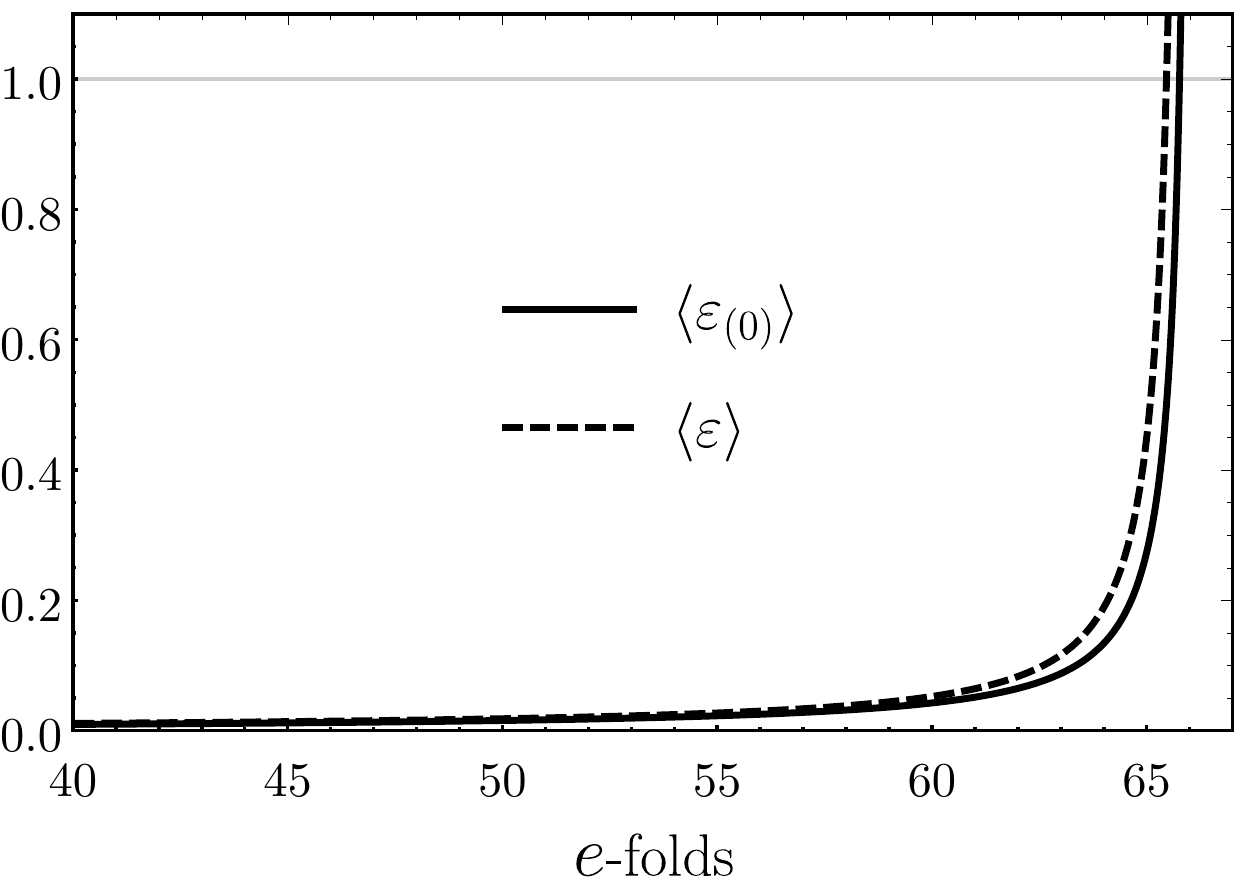}
\includegraphics[width=0.45\textwidth]{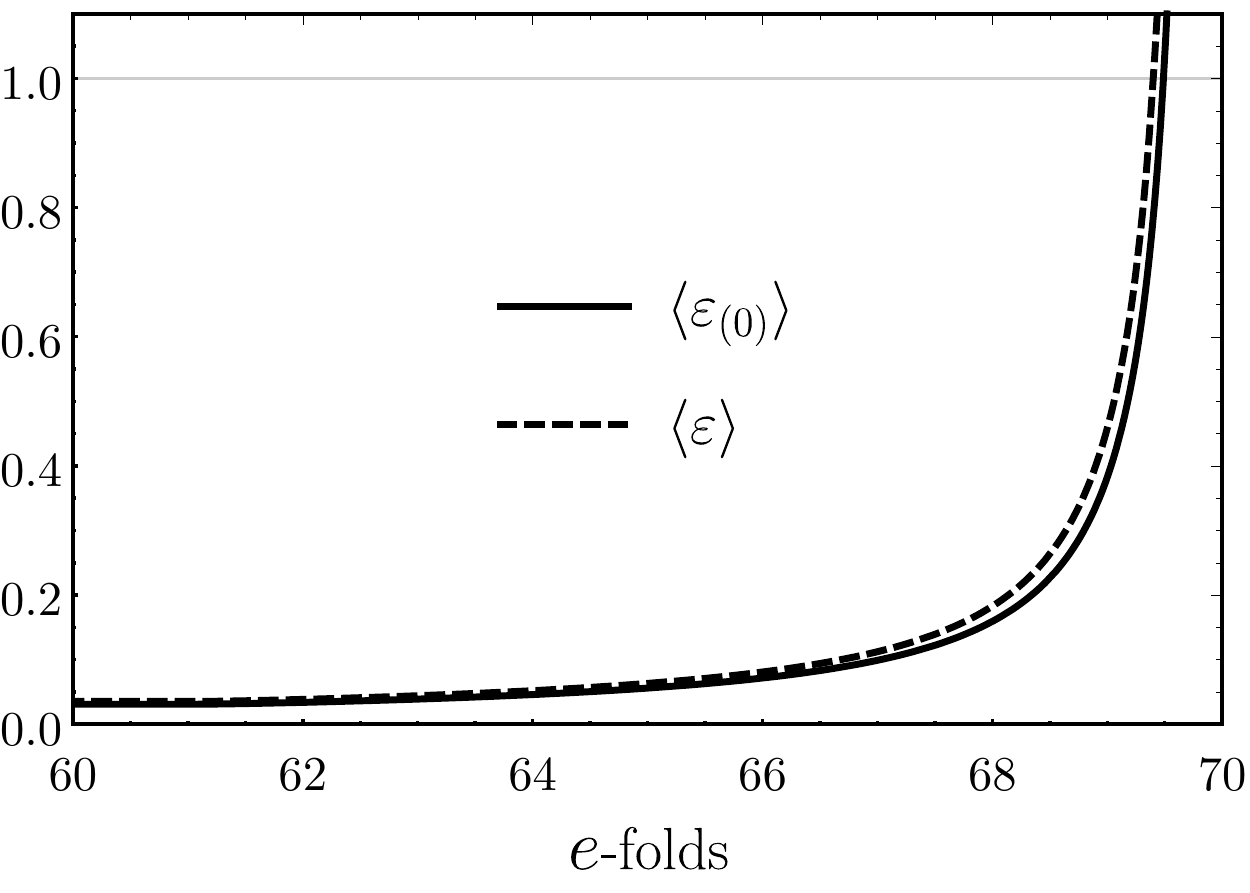}
\caption{Evolution of $\varepsilon$ for set 1 (left) and set 2 (right).  The quantity $\langle \varepsilon_{(0)}\rangle $ represents the first slow roll parameter averaged over oscillations in the potential and without the contribution due to the backreaction on the K\"ahler moduli, while $\langle \varepsilon\rangle $ is the averaged slow roll parameter taking backreaction into account. }\label{fig-epsilon}
\end{center}
\end{figure}

\begin{figure}[H]
\begin{center}
\includegraphics[width=0.45\textwidth]{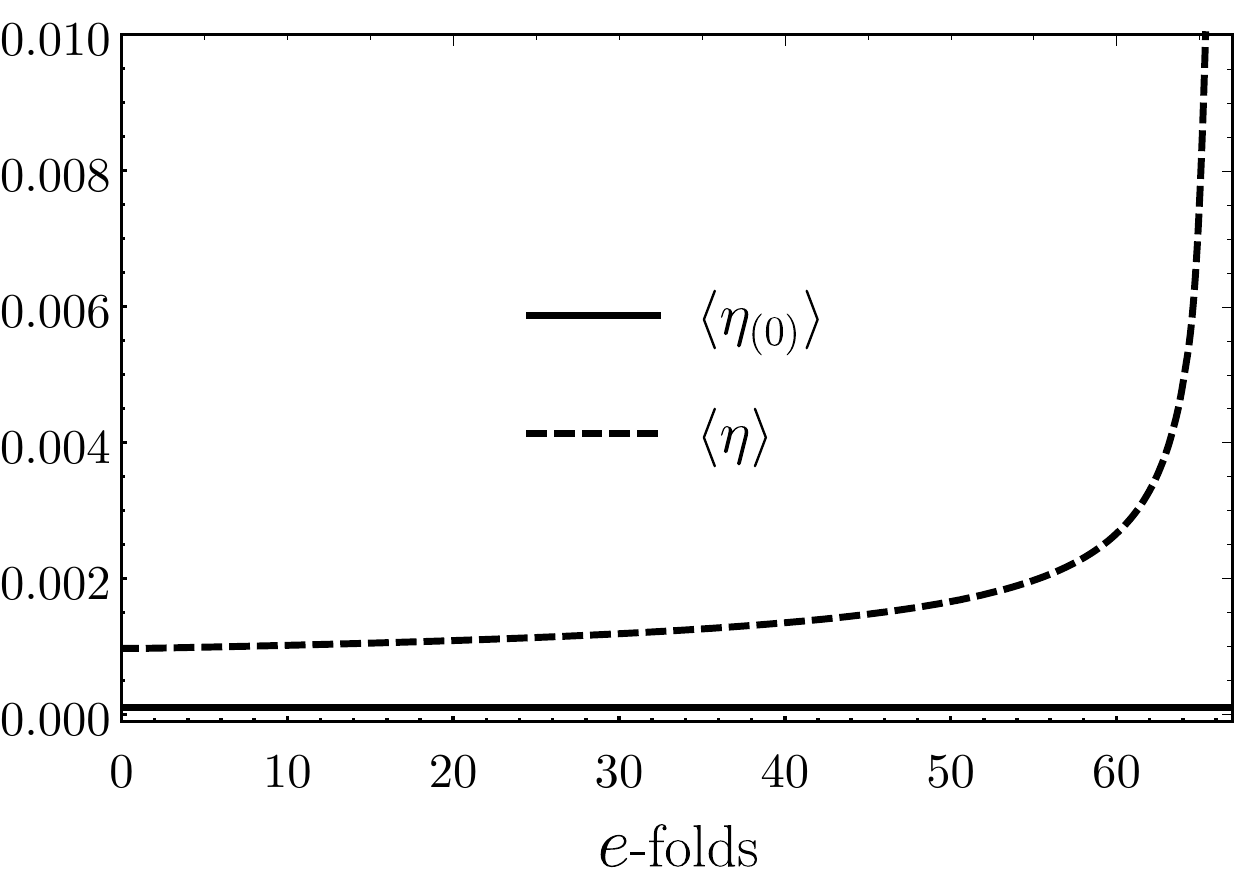}
\includegraphics[width=0.45\textwidth]{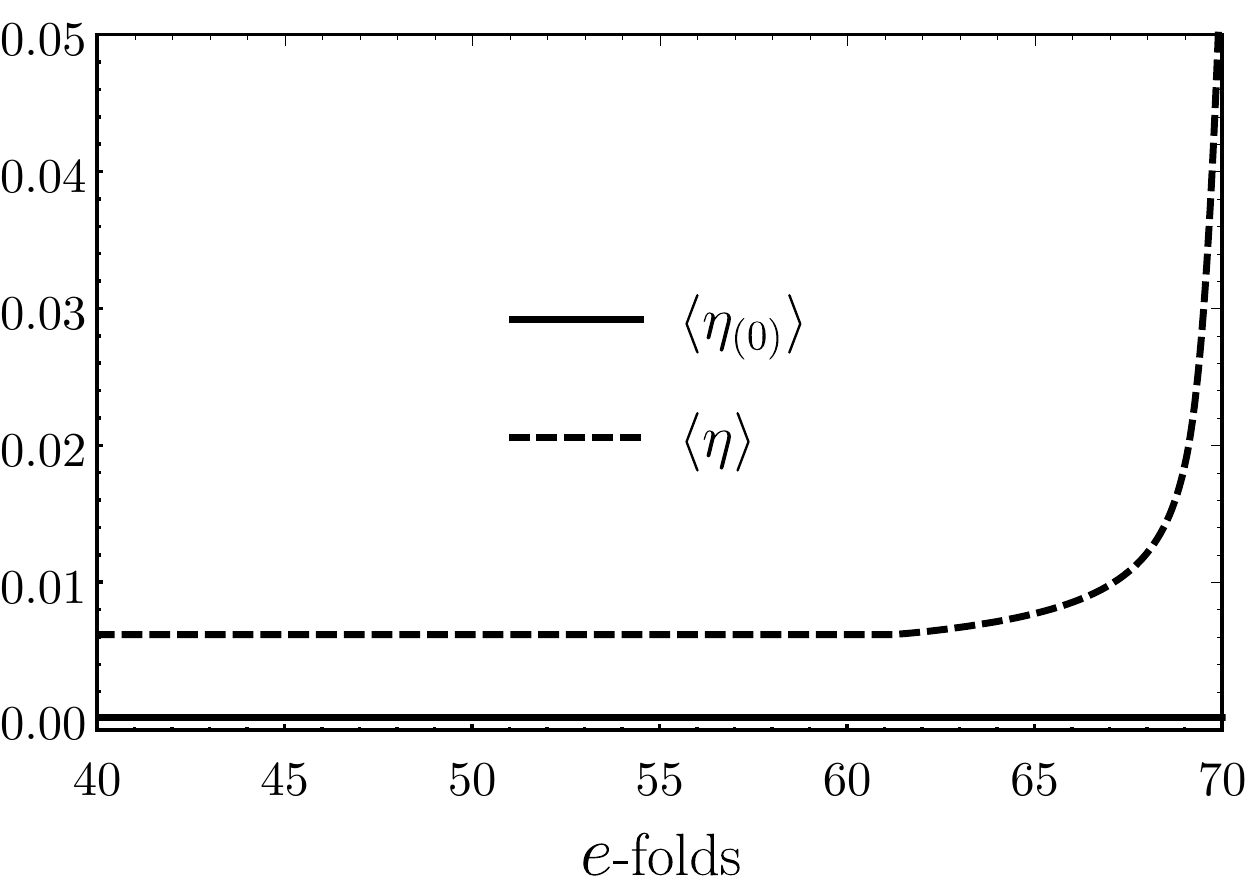}
\caption{Evolution of $\eta$ for set 1 (left) and set 2 (right).  While the averaged $\eta$ parameter is zero in the absence of backreaction, including the dependence of the K\"ahler moduli on the inflaton gives rise to a small but non-zero average.}\label{fig-eta}
\end{center}
\end{figure}

In order to compute the backreaction effects we need functions $\V(\phi)$ and $\tau(\phi)$, which  are computed by numerically solving for the minimum of \eqref{fullKaehlerPot} at each point in the cascade.  Because the K\"ahler potential \eqref{fullKaehlerPot} is rather messy, we will not bother writing the analytic expressions for the slow roll parameters, but rather refer to Figures \ref{fig-epsilon} and \ref{fig-eta} for the results.  We find that while the backreaction indeed increases $\langle \eta\rangle >0$, it remains true that   $\langle \eta\rangle ,\langle \varepsilon\rangle \ll 1$ until the end of inflation, which we define by\footnote{Strictly speaking we use the slow roll parameter $\varepsilon_H = \dot{H}/H^2$ to define the end of inflation as it additionally accounts for changes in the field's kinetic energy.} $\langle \varepsilon\rangle =1$.  Here, the angle brackets indicate averaging over oscillations in the potential.  While $\varepsilon$ was found to be small for all scenarios with $p<MK$, the small contribution to $\langle \eta\rangle$ is far less trivial and considerably tightened the constraints on parameter space.

In addition to potentially dangerous contributions to the potential, it is also important to calculate the effect of backreaction on the field excursion, which could spoil the claim of a trans-Planckian field range.  Specifically, the $\V$ dependence in \eqref{canfieldSDKS} should not be treated as constant and the canonical field should be corrected:
\be \label{backreactfield}
\d \phi = \tilde{f}(\V(\psi))\d \psi~.
\ee
Taking this backreaction into account we indeed find that the field range is decreased, although not dramatically when the $\eta$-problem is avoided.  Actually,  the fact that $\varepsilon$ is slightly larger when backreaction is included, resulting in a slightly earlier end of inflation makes a larger impact on the field range than the correction \eqref{backreactfield}. The backreacted field ranges are collected in Table \ref{table-bkrxn}. 

\subsection{Breaking supersymmetry} \label{bkrxnnosusy}
The backreaction of antibranes on the KS geometry is a vast topic.  Particularly, the backreaction in the deep IR -- at the tip of the throat -- has been the subject of longstanding debates. Both point-like and smeared anti-D3 brane sources have been shown to give rise to ill-behaved singularities \cite{McGuirk:2009xx,Bena:2009xk,Bena:2011hz,Bena:2011wh}. Recently, \cite{Cohen-Maldonado:2015ssa} showed (building upon \cite{Gautason:2013zw,Blaback:2014tfa}) that localized NS5-branes carrying anti-D3 charge avoid previous no-go arguments given against the existence of regular IR boundary conditions, and therefore provide a viable candidate for a regular supersymmetry breaking solution\footnote{See also \cite{Michel:2014lva,Hartnett:2015oda}.}. In this work we will only consider the supersymmetry breaking backreaction in the UV, which should be enough for the purposes of the backreaction on the warped volume of the four-cycle supporting non-perturbative effects, and the sub-leading backreaction of the warp factor on the moduli.

The asymptotic analysis of \cite{DeWolfe:2008zy}, together with the amendments of \cite{Bena:2009xk,Bena:2011hz,Bena:2011wh} and further insights of \cite{Dymarsky:2011pm,Dymarsky:2013tna}, finds that antibranes at the tip of the KS throat break supersymmetry, squash the KS geometry and induce a radial running on the dilaton. A full calculation of backreaction for this model would include calculating the warped volume of the four-cycle relevant for non-perturbative effects using the backreacted metric, which depends on the antibrane charge and therefore on our inflaton.  We will merely dip our toes into this full calculation  by noting that the strength of the perturbations to the KS geometry and running of the dilaton is expected to be proportional to \cite{DeWolfe:2008zy}\footnote{The holographic interpretation of this parameter was recently clarified in \cite{Bertolini:2015hua}.}:
\begin{align}
{\cal S} &\sim \frac{p}{KM-p}\e^{\frac{-8\pi(KM-p)}{3g_s M^2}}\qquad \rm{KS}~, \label{SKS}\\
\tilde{{\cal S}} &\sim \frac{p}{KM-p}\e^{\frac{-8\pi g_s (KM-p)}{3 K^2}}\qquad \rm{SDKS}~. \label{SSDKS}
\end{align}
Very schematically, the warp factor is affected as $h\to h(1+{\cal S}/r^4)$, and similarly for the dilaton. Thus the change to the metric in the UV should be small as long as these quantities can be tuned to be small. Furthermore, the inflaton dependence will arise only linearly via $p(\phi)$ in \eqref{SKS} and \eqref{SSDKS}  and with a strong exponential suppression (again, the exponent is inflaton independent due to charge conservation.)  We collect a summary of the effects of backreaction and the size of the supersymmetry breaking effects in Table \ref{table-bkrxn}.   
\begin{table}[H]
\begin{center}
\begin{tabular}{|c|c|c|c|}
\hline 
• & $\Delta \phi_{bk}/M_p$ & $\langle \eta \rangle|_{t*}$ & $\tilde{\cal S}$ \\ 
\hline 
Set 1 & 10.4 & $1.0\cdot 10^{-3}$ & $2.2\cdot 10^{-12}$ \\ 
\hline 
Set 2 & 3.52 & $6.3\cdot10^{-3}$& $1.7\cdot 10^{-13}$ \\ 
\hline 
\end{tabular}
\end{center}
\caption{The effects of backreaction for the two parameter sets given in \secref{Examples}. The averaged second slow roll parameter is evaluated at the observationally relevant window, 60 efolds before the end of inflation, and will alter the value of $n_s$ given in Table \ref{table-field+pspec} }\label{table-bkrxn}
\end{table}

\section{Conclusions} \label{Conclusions}
In this work we have taken steps towards an explicit model of large field inflation.  We achieve an inflationary epoch with a super-Planckian field excursion by placing a stack of anti-D3 branes in a throat geometry.  The use of simple and well-studied ingredients, such as anti-D3 branes and throat geometries in flux compactifications, make this one of the most tractable examples of large field inflation in string theory.  Thus, we were able to compute leading backreaction effects and find scenarios for which neither the inflationary epoch nor the field excursion is strongly affected by the backreaction on K\"ahler moduli.  

These promising results point to avenues for future research.  First, in order to move towards a fully explicit model, we will need a Calabi-Yau three-fold which contains a warped throat region and either 1) contains large negative sources of D3 charge outside the unwinding throat, or 2) lifts to a Calabi-Yau four-fold in F-theory that has a very large Euler number.  The difficulty in finding such an example could place strong constraints on possible scenarios.  Second, even without a fully explicit manifold, additional UV backreaction effects beyond the s-wave approximation and taking into account the effects of supersymmetry breaking discussed in \secref{bkrxnnosusy} are straightforward and tractable calculations.  Meanwhile, a full analysis beyond the probe approximation in the IR remains a challenging but interesting topic for future research.

As opposed to computing inflationary observables with many significant figures, we would like to highlight  the relative simplicity of this model as an ideal setting to further investigate the important questions of what is allowed in the string landscape.  Furthermore, while it is clear that the string cosmology community would benefit from a better understanding of the basic ingredients necessary to find de Sitter phases in string theory, strong observational evidence suggests that current technical barriers to rigorously achieving an inflationary epoch will be overcome.  Therefore, it is important to understand simple and flexible inflationary scenarios which can adapt to our evolving understanding of positive vacuum energies in string theory.  We hope that antibranes, brane flux annihilation, and warped throats will prove robust enough that unwinding inflation can continue to be relevant in the study of string cosmology.

\section*{Acknowledgements}
We would like to thank Thomas Bachlachner, Fridrik Freyr Gautason, Oliver Janssen, Matthew Kleban, Liam McAllister, Thomas Van Riet and Bert Vercnocke for useful discussions.   Additionally, we thank  Fridrik Freyr Gautason and Thomas Van Riet for detailed feedback on the manuscript.  JDD is supported by the National Science Foundation of Belgium (FWO) grant G.0.E52.14N Odysseus. MS is supported by the European Union's Horizon 2020
research and innovation programme under the Marie Sk\l odowska-Curie grant agreement No.\,656491.

\appendix
\section{Examples in Klebanov-Strassler} \label{app-KS}
We present here two inflationary sets in KS with roughly the same characteristics as the sets discussed in the main text.
\begin{table}[H]
\begin{center}
\begin{tabular}{|c|c|c|c|c|c|c|c|c|c|c|}
\hline 
• & $a$ & $\xi$ & $A$ & $W_0$ & $g_s$ & $M$ & $K$ & $p$ & $\tau$ & $\cal{V}$  \\ 
\hline 
Set 1 & $\pi$ & 1.22 & 1 & $2\cdot10^6$ & .248 & 19\,650 & 21\,010 & 41\,284\,650 & .772 & $1.73\cdot 10^9$  \\ 
\hline 
Set 2 & $\pi$ & 1.22 & 1 & $2\cdot10^6$ & .248 & 18\,870 & 18\,010 & 11\,328\,290 & .773 & $1.74\cdot 10^9$ \\ 
\hline 
\end{tabular}
\end{center}
\caption{Two sets of parameters which give rise to two different inflationary regimes in KS.}\label{table-param-KS}
\end{table}

\begin{table}[H]
\begin{center}
\begin{tabular}{|c|c|c|c|c|c|}
\hline 
• & $e$-folds & $\Delta\phi/M_P$ & $\mathcal{P}_\zeta$ & $n_s$ & $r$ \\ 
\hline 
Set 1 & 65 & 10.8 & 7.84$\cdot10^{-22}$ & .978 & .050 \\ 
\hline 
Set 2 & 70 & 2.04 & 2.84$\cdot 10^{-9}$ & 1.00 & 4.29$\cdot10^{-14}$ \\ 
\hline 
\end{tabular}
\end{center}
\caption{Field range and power spectrum for both parameter sets. }\label{table-field+pspec-KS}
\end{table}

\begin{table}[H]
\begin{center}
\begin{tabular}{|c|c|c|c|c|c|c|c|c|}
\hline 
• & $\frac{p}{KM}$ & $\frac{p}{g_sM^2}$ & $e^{-a\tau/g_s}$ & $\frac{\mathcal{V}^{2/3}}{h_0}$ & $\frac{\mathcal{V}_{throat}}{\mathcal{V}}$ & $\mathcal{V}H^2$ & $\frac{H^2}{m_{\overline{\tau}}^2}$ & $\frac{H^2}{m_{\overline{\tau}_b}^2}$ \\ 
\hline 
Set 1 & .100 & .432 & 5.52$\cdot10^{-5}$ & 7.45$\cdot10^{-7}$ & .294 & 6.42$\cdot10^{-3}$ & 7.12$\cdot10^{-7}$ & .694  \\ 
\hline 
Set 2 & 0.033 & 0.129 & 5.49$\cdot10^{-5}$ & 2.74$\cdot10^{-6}$ & .241 & 7.01$\cdot10^{-3}$ & 7.43$\cdot10^{-7}$ & .727  \\ 
\hline 
\end{tabular}
\end{center}
\caption{All the constraints mentioned in \secref{constraints} are satisfied by both parameter sets.  }\label{table-constraints-KS}
\end{table}

\begin{table}[H]
\begin{center}
\begin{tabular}{|c|c|c|c|}
\hline 
• & $\Delta \phi_{bk}/M_p$ & $\langle \eta \rangle|_{t*}$ & $\cal S$ \\ 
\hline 
Set 1 & 10.1 & $2.1\cdot 10^{-3}$ & $7.9\cdot 10^{-16}$ \\ 
\hline 
Set 2 & 1.96 & $5.0\cdot10^{-2}$& $9.4\cdot 10^{-16}$ \\ 
\hline 
\end{tabular}
\end{center}
\caption{The effects of backreaction for the two parameter sets.}\label{table-bkrxn-KS}
\end{table}

\begin{figure}[H]
\begin{center}
\includegraphics[width=0.46\textwidth]{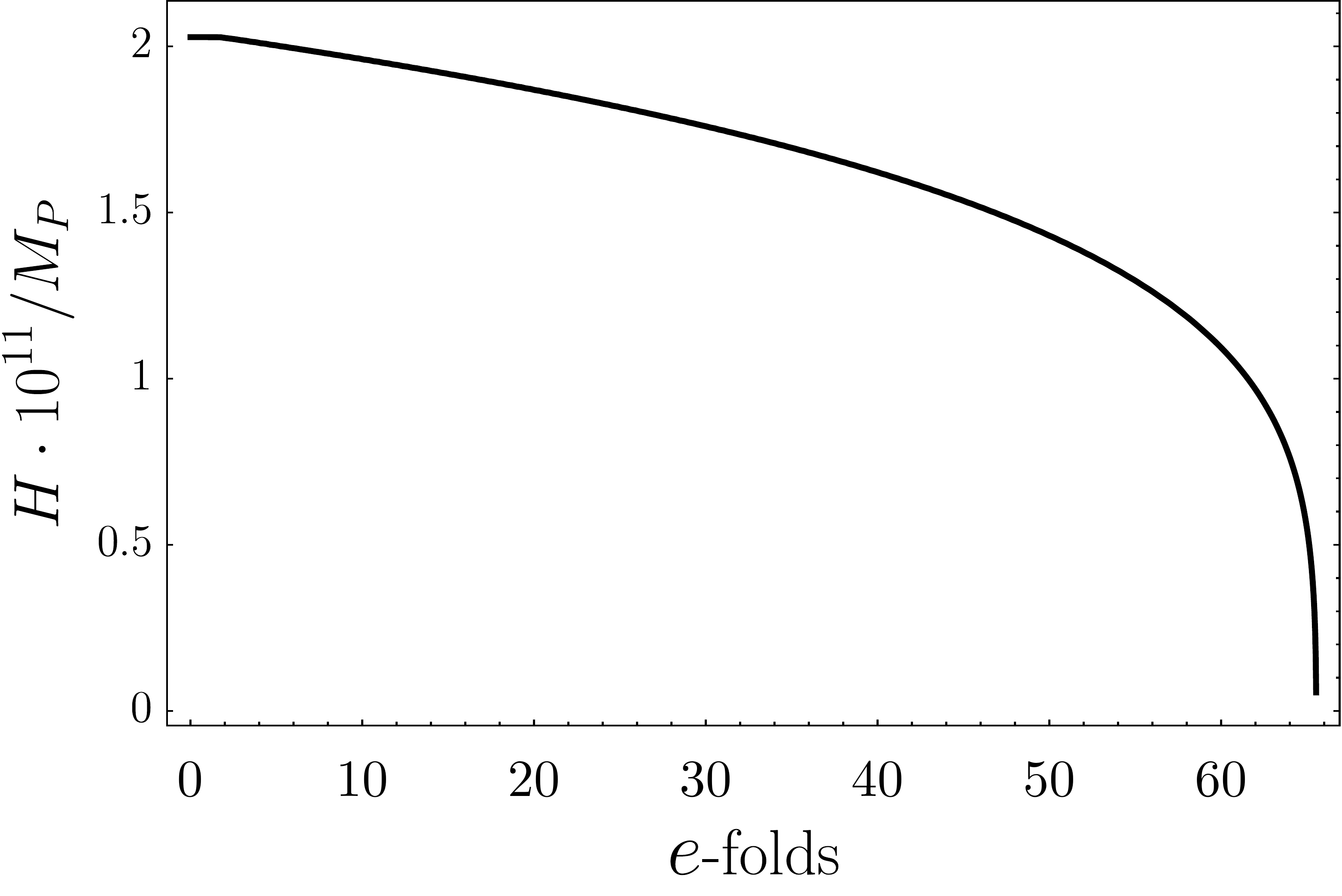}
\includegraphics[width=0.46\textwidth]{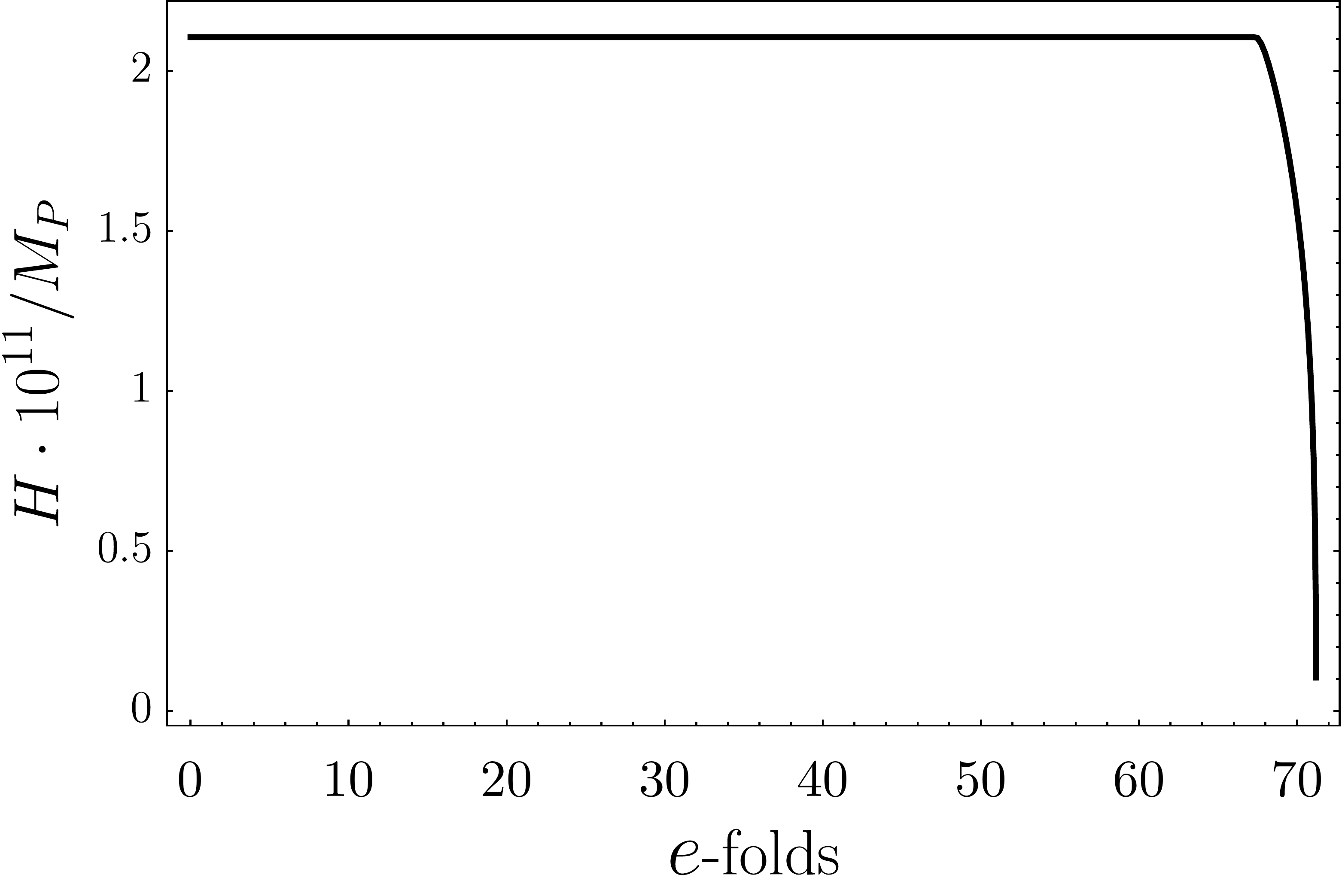}
\caption{Evolution of the Hubble parameter for set 1 (left) and set 2 (right).}\label{fig-Hubble-KS}
\end{center}
\end{figure}

\begin{figure}[H]
\begin{center}
\includegraphics[width=0.45\textwidth]{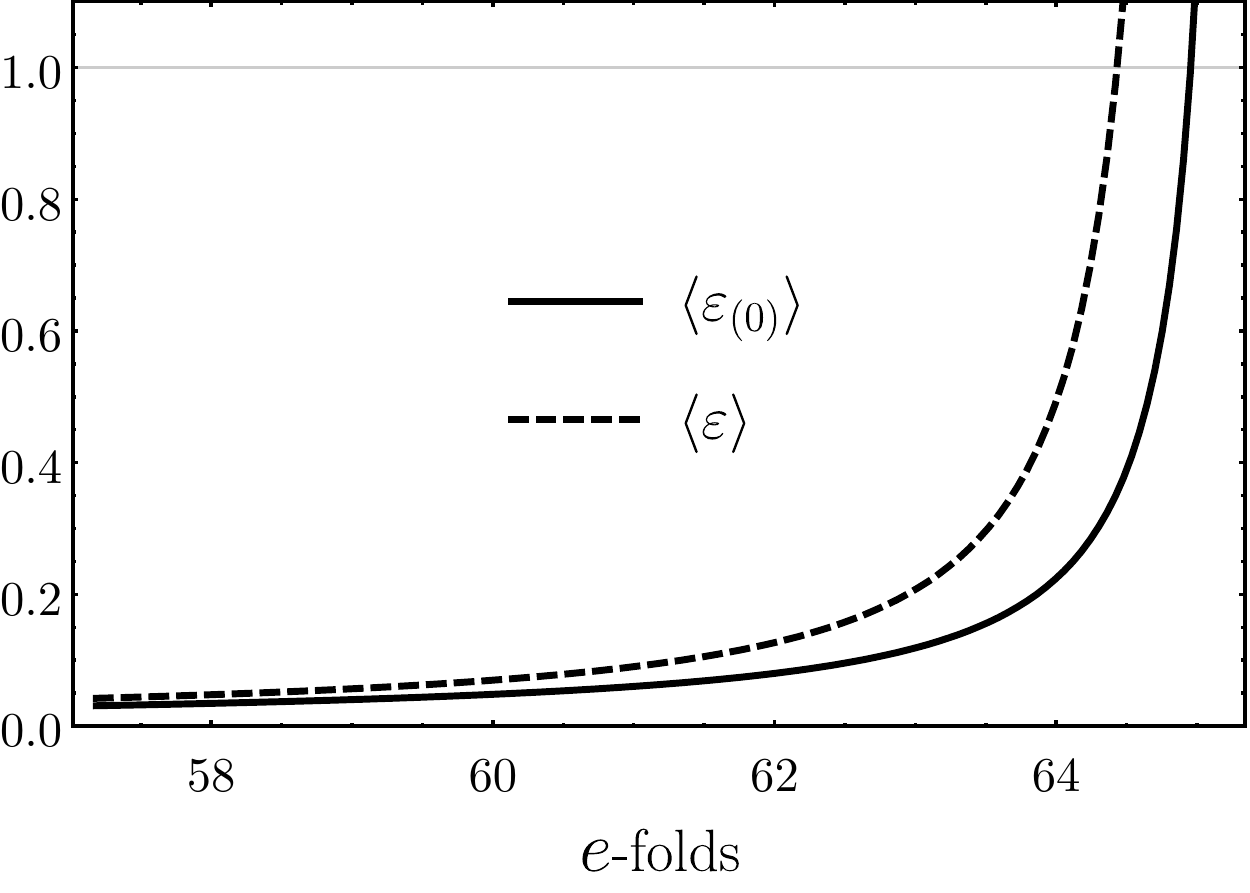}
\includegraphics[width=0.45\textwidth]{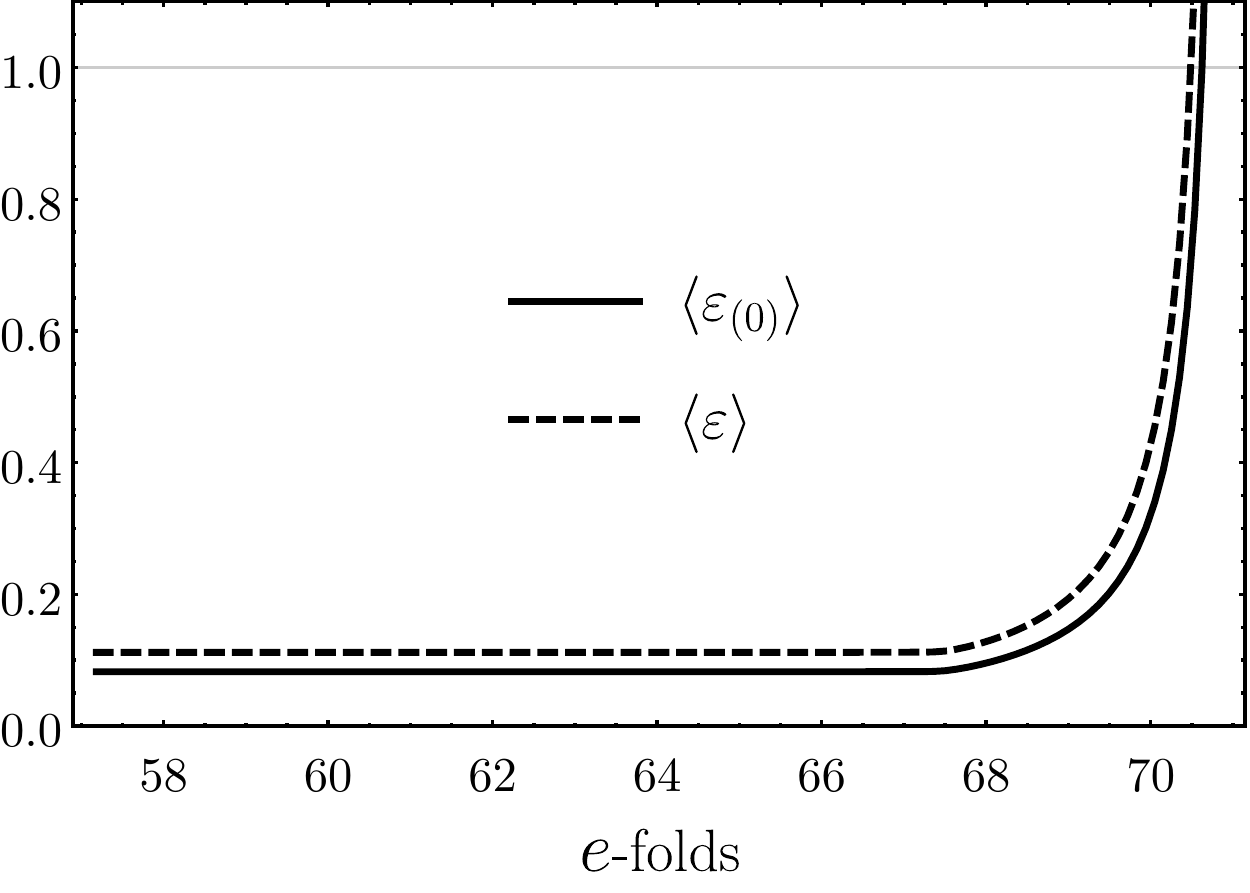}
\caption{Evolution of $\varepsilon$ for set 1 (left) and set 2 (right).}\label{fig-epsilon-KS}
\end{center}
\end{figure}

\begin{figure}[H]
\begin{center}
\includegraphics[width=0.45\textwidth]{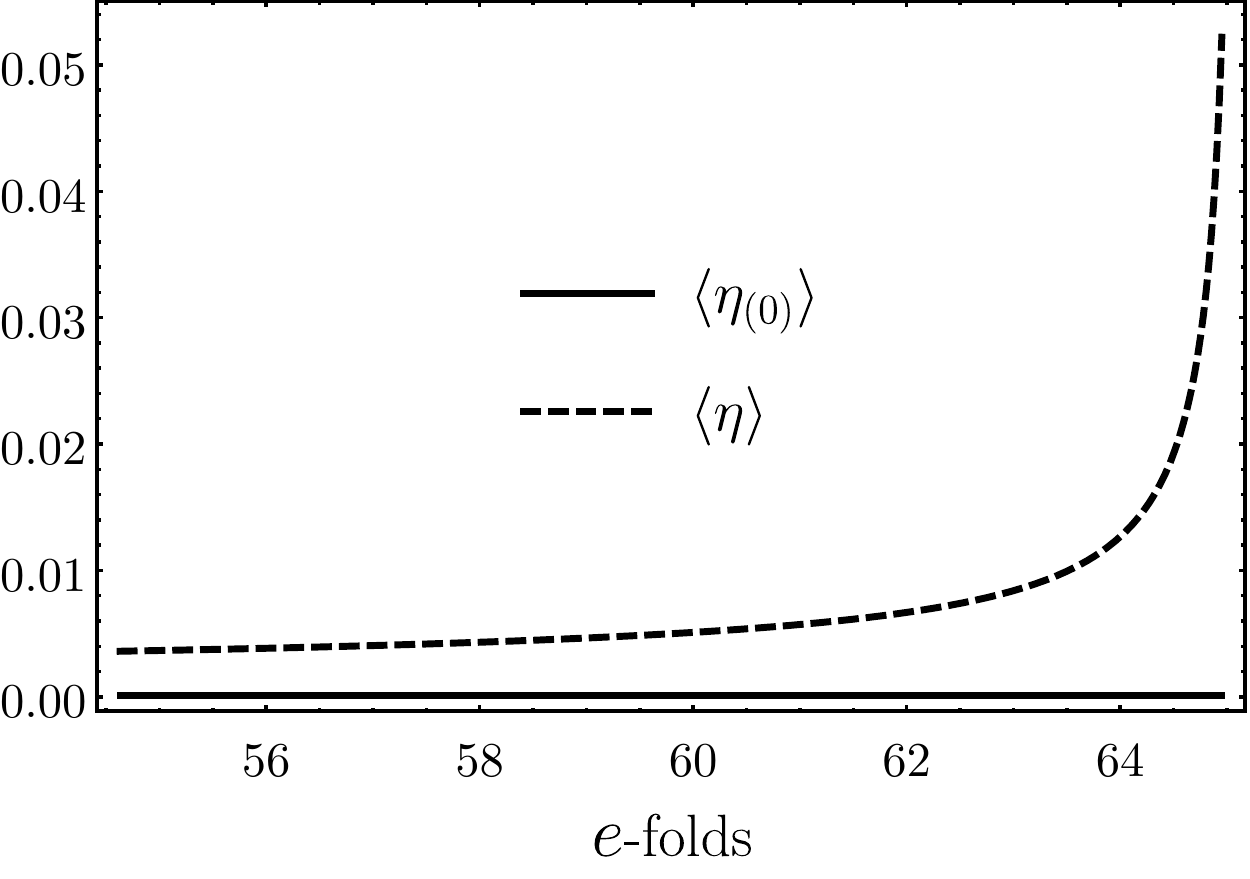}
\includegraphics[width=0.45\textwidth]{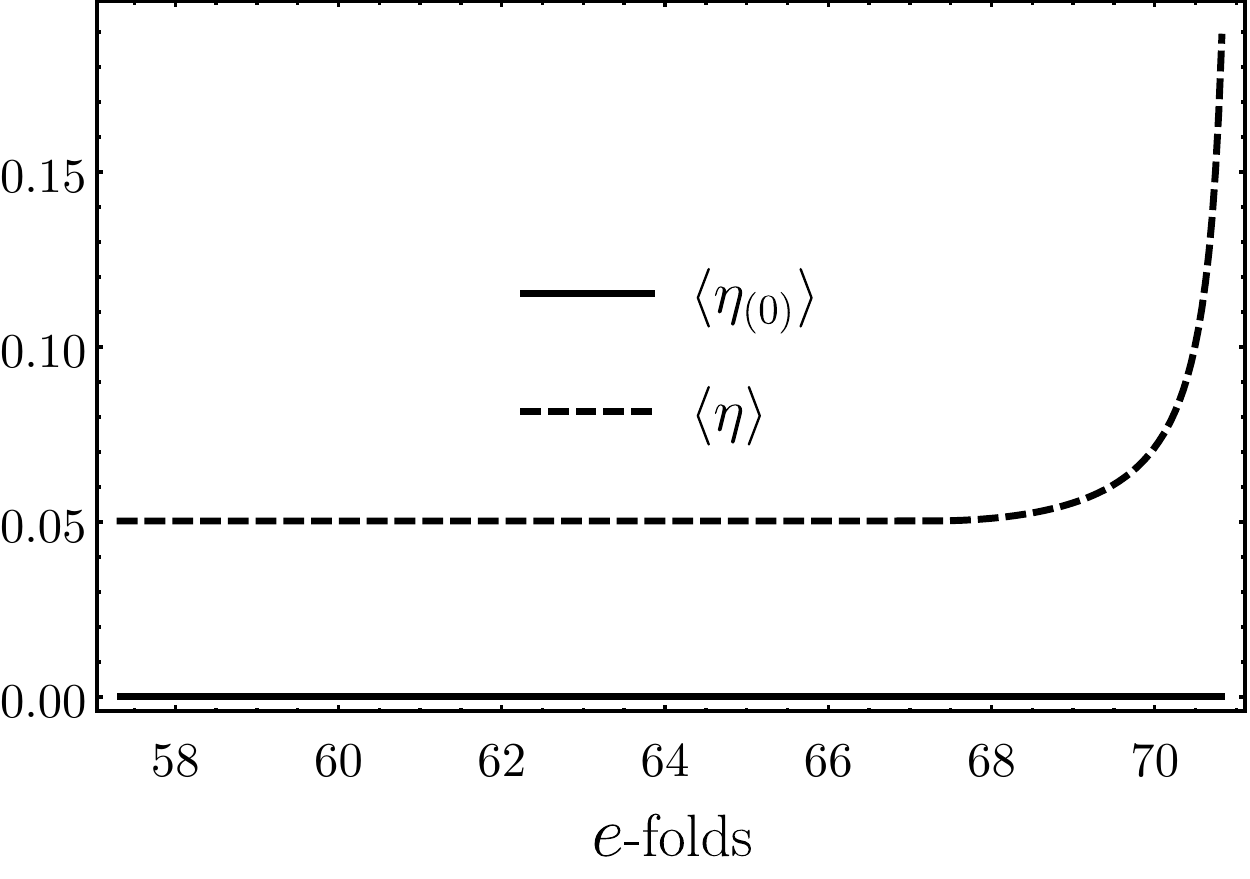}
\caption{Evolution of $\eta$ for set 1 (left) and set 2 (right).}\label{fig-eta-KS}
\end{center}
\end{figure}

\bibliography{refs}

\providecommand{\href}[2]{#2}\begingroup\raggedright\begin{thebibliography}{10}

\bibitem{Ade:2015lrj}
{\bfseries Planck} Collaboration, {\sc P.~A.~R. Ade} et~al., ``{Planck 2015
  results. XX. Constraints on inflation},''
  \href{http://dx.doi.org/10.1051/0004-6361/201525898}{{\em Astron. Astrophys.}
  {\bfseries 594} (2016) A20},
\href{http://arxiv.org/abs/1502.02114}{{\ttfamily arXiv:1502.02114
  [astro-ph.CO]}}.

\bibitem{Vafa:2005ui}
{\sc C.~Vafa}, ``{The String landscape and the swampland},''
\href{http://arxiv.org/abs/hep-th/0509212}{{\ttfamily arXiv:hep-th/0509212
  [hep-th]}}.

\bibitem{Ooguri:2006in}
{\sc H.~Ooguri} and {\sc C.~Vafa}, ``{On the Geometry of the String Landscape
  and the Swampland},''
  \href{http://dx.doi.org/10.1016/j.nuclphysb.2006.10.033}{{\em Nucl. Phys.}
  {\bfseries B766} (2007) 21--33},
\href{http://arxiv.org/abs/hep-th/0605264}{{\ttfamily arXiv:hep-th/0605264
  [hep-th]}}.

\bibitem{ArkaniHamed:2006dz}
{\sc N.~Arkani-Hamed}, {\sc L.~Motl}, {\sc A.~Nicolis}, and {\sc C.~Vafa},
  ``{The String landscape, black holes and gravity as the weakest force},''
  \href{http://dx.doi.org/10.1088/1126-6708/2007/06/060}{{\em JHEP} {\bfseries
  06} (2007) 060},
\href{http://arxiv.org/abs/hep-th/0601001}{{\ttfamily arXiv:hep-th/0601001
  [hep-th]}}.

\bibitem{Brennan:2017rbf}
{\sc T.~D. Brennan}, {\sc F.~Carta}, and {\sc C.~Vafa}, ``{The String
  Landscape, the Swampland, and the Missing Corner},''
\href{http://arxiv.org/abs/1711.00864}{{\ttfamily arXiv:1711.00864 [hep-th]}}.

\bibitem{D'Amico:2012ji}
{\sc G.~D'Amico}, {\sc R.~Gobbetti}, {\sc M.~Kleban}, and {\sc M.~Schillo},
  ``{Unwinding Inflation},''
  \href{http://dx.doi.org/10.1088/1475-7516/2013/03/004}{{\em JCAP} {\bfseries
  1303} (2013) 004},
\href{http://arxiv.org/abs/1211.4589}{{\ttfamily arXiv:1211.4589 [hep-th]}}.

\bibitem{Gautason:2016cyp}
{\sc F.~F. Gautason}, {\sc M.~Schillo}, and {\sc T.~Van~Riet}, ``{Is inflation
  from unwinding fluxes IIB?},''
  \href{http://dx.doi.org/10.1007/JHEP03(2017)037}{{\em JHEP} {\bfseries 03}
  (2017) 037},
\href{http://arxiv.org/abs/1611.07037}{{\ttfamily arXiv:1611.07037 [hep-th]}}.

\bibitem{Klebanov:2000hb}
{\sc I.~R. Klebanov} and {\sc M.~J. Strassler}, ``{Supergravity and a confining
  gauge theory: Duality cascades and chi SB resolution of naked
  singularities},'' \href{http://dx.doi.org/10.1088/1126-6708/2000/08/052}{{\em
  JHEP} {\bfseries 08} (2000) 052},
\href{http://arxiv.org/abs/hep-th/0007191}{{\ttfamily arXiv:hep-th/0007191
  [hep-th]}}.

\bibitem{Giddings:2001yu}
{\sc S.~B. Giddings}, {\sc S.~Kachru}, and {\sc J.~Polchinski}, ``{Hierarchies
  from fluxes in string compactifications},''
  \href{http://dx.doi.org/10.1103/PhysRevD.66.106006}{{\em Phys. Rev.}
  {\bfseries D66} (2002) 106006},
\href{http://arxiv.org/abs/hep-th/0105097}{{\ttfamily arXiv:hep-th/0105097
  [hep-th]}}.

\bibitem{Balasubramanian:2005zx}
{\sc V.~Balasubramanian}, {\sc P.~Berglund}, {\sc J.~P. Conlon}, and {\sc
  F.~Quevedo}, ``{Systematics of moduli stabilisation in Calabi-Yau flux
  compactifications},''
  \href{http://dx.doi.org/10.1088/1126-6708/2005/03/007}{{\em JHEP} {\bfseries
  03} (2005) 007},
\href{http://arxiv.org/abs/hep-th/0502058}{{\ttfamily arXiv:hep-th/0502058
  [hep-th]}}.

\bibitem{Conlon:2005ki}
{\sc J.~P. Conlon}, {\sc F.~Quevedo}, and {\sc K.~Suruliz}, ``{Large-volume
  flux compactifications: Moduli spectrum and D3/D7 soft supersymmetry
  breaking},'' \href{http://dx.doi.org/10.1088/1126-6708/2005/08/007}{{\em
  JHEP} {\bfseries 08} (2005) 007},
\href{http://arxiv.org/abs/hep-th/0505076}{{\ttfamily arXiv:hep-th/0505076
  [hep-th]}}.

\bibitem{Kachru:2002gs}
{\sc S.~Kachru}, {\sc J.~Pearson}, and {\sc H.~L. Verlinde}, ``{Brane / flux
  annihilation and the string dual of a nonsupersymmetric field theory},''
  \href{http://dx.doi.org/10.1088/1126-6708/2002/06/021}{{\em JHEP} {\bfseries
  06} (2002) 021},
\href{http://arxiv.org/abs/hep-th/0112197}{{\ttfamily arXiv:hep-th/0112197
  [hep-th]}}.

\bibitem{Kleban:2011cs}
{\sc M.~Kleban}, {\sc K.~Krishnaiyengar}, and {\sc M.~Porrati}, ``{Flux
  Discharge Cascades in Various Dimensions},''
  \href{http://dx.doi.org/10.1007/JHEP11(2011)096}{{\em JHEP} {\bfseries 11}
  (2011) 096},
\href{http://arxiv.org/abs/1108.6102}{{\ttfamily arXiv:1108.6102 [hep-th]}}.

\bibitem{DeWolfe:2008zy}
{\sc O.~DeWolfe}, {\sc S.~Kachru}, and {\sc M.~Mulligan}, ``{A Gravity Dual of
  Metastable Dynamical Supersymmetry Breaking},''
  \href{http://dx.doi.org/10.1103/PhysRevD.77.065011}{{\em Phys. Rev.}
  {\bfseries D77} (2008) 065011},
\href{http://arxiv.org/abs/0801.1520}{{\ttfamily arXiv:0801.1520 [hep-th]}}.

\bibitem{Bena:2009xk}
{\sc I.~Bena}, {\sc M.~Grana}, and {\sc N.~Halmagyi}, ``{On the Existence of
  Meta-stable Vacua in Klebanov-Strassler},''
  \href{http://dx.doi.org/10.1007/JHEP09(2010)087}{{\em JHEP} {\bfseries 09}
  (2010) 087},
\href{http://arxiv.org/abs/0912.3519}{{\ttfamily arXiv:0912.3519 [hep-th]}}.

\bibitem{Bena:2011hz}
{\sc I.~Bena}, {\sc G.~Giecold}, {\sc M.~Grana}, {\sc N.~Halmagyi}, and {\sc
  S.~Massai}, ``{On Metastable Vacua and the Warped Deformed Conifold: Analytic
  Results},'' \href{http://dx.doi.org/10.1088/0264-9381/30/1/015003}{{\em
  Class. Quant. Grav.} {\bfseries 30} (2013) 015003},
\href{http://arxiv.org/abs/1102.2403}{{\ttfamily arXiv:1102.2403 [hep-th]}}.

\bibitem{Bena:2011wh}
{\sc I.~Bena}, {\sc G.~Giecold}, {\sc M.~Grana}, {\sc N.~Halmagyi}, and {\sc
  S.~Massai}, ``{The backreaction of anti-D3 branes on the Klebanov-Strassler
  geometry},'' \href{http://dx.doi.org/10.1007/JHEP06(2013)060}{{\em JHEP}
  {\bfseries 06} (2013) 060},
\href{http://arxiv.org/abs/1106.6165}{{\ttfamily arXiv:1106.6165 [hep-th]}}.

\bibitem{Dymarsky:2011pm}
{\sc A.~Dymarsky}, ``{On gravity dual of a metastable vacuum in
  Klebanov-Strassler theory},''
  \href{http://dx.doi.org/10.1007/JHEP05(2011)053}{{\em JHEP} {\bfseries 05}
  (2011) 053},
\href{http://arxiv.org/abs/1102.1734}{{\ttfamily arXiv:1102.1734 [hep-th]}}.

\bibitem{Dymarsky:2013tna}
{\sc A.~Dymarsky} and {\sc S.~Massai}, ``{Uplifting the baryonic branch: a test
  for backreacting anti-D3-branes},''
  \href{http://dx.doi.org/10.1007/JHEP11(2014)034}{{\em JHEP} {\bfseries 11}
  (2014) 034},
\href{http://arxiv.org/abs/1310.0015}{{\ttfamily arXiv:1310.0015 [hep-th]}}.

\bibitem{Moritz:2017xto}
{\sc J.~Moritz}, {\sc A.~Retolaza}, and {\sc A.~Westphal}, ``{Towards de Sitter
  from 10D},''
\href{http://arxiv.org/abs/1707.08678}{{\ttfamily arXiv:1707.08678 [hep-th]}}.

\bibitem{Kachru:2003aw}
{\sc S.~Kachru}, {\sc R.~Kallosh}, {\sc A.~D. Linde}, and {\sc S.~P. Trivedi},
  ``{De Sitter vacua in string theory},''
  \href{http://dx.doi.org/10.1103/PhysRevD.68.046005}{{\em Phys. Rev.}
  {\bfseries D68} (2003) 046005},
\href{http://arxiv.org/abs/hep-th/0301240}{{\ttfamily arXiv:hep-th/0301240
  [hep-th]}}.

\bibitem{Sethi:2017phn}
{\sc S.~Sethi}, ``{Supersymmetry Breaking by Fluxes},''
\href{http://arxiv.org/abs/1709.03554}{{\ttfamily arXiv:1709.03554 [hep-th]}}.

\bibitem{Gukov:1999ya}
{\sc S.~Gukov}, {\sc C.~Vafa}, and {\sc E.~Witten}, ``{CFT's from Calabi-Yau
  four folds},'' \href{http://dx.doi.org/10.1016/S0550-3213(01)00289-9,
  10.1016/S0550-3213(00)00373-4}{{\em Nucl. Phys.} {\bfseries B584} (2000)
  69--108}, \href{http://arxiv.org/abs/hep-th/9906070}{{\ttfamily
  arXiv:hep-th/9906070 [hep-th]}}.
[Erratum: Nucl. Phys.B608,477(2001)].

\bibitem{Becker:2002nn}
{\sc K.~Becker}, {\sc M.~Becker}, {\sc M.~Haack}, and {\sc J.~Louis},
  ``{Supersymmetry breaking and alpha-prime corrections to flux induced
  potentials},'' \href{http://dx.doi.org/10.1088/1126-6708/2002/06/060}{{\em
  JHEP} {\bfseries 06} (2002) 060},
\href{http://arxiv.org/abs/hep-th/0204254}{{\ttfamily arXiv:hep-th/0204254
  [hep-th]}}.

\bibitem{Giddings:2005ff}
{\sc S.~B. Giddings} and {\sc A.~Maharana}, ``{Dynamics of warped
  compactifications and the shape of the warped landscape},''
  \href{http://dx.doi.org/10.1103/PhysRevD.73.126003}{{\em Phys. Rev.}
  {\bfseries D73} (2006) 126003},
\href{http://arxiv.org/abs/hep-th/0507158}{{\ttfamily arXiv:hep-th/0507158
  [hep-th]}}.

\bibitem{Frey:2008xw}
{\sc A.~R. Frey}, {\sc G.~Torroba}, {\sc B.~Underwood}, and {\sc M.~R.
  Douglas}, ``{The Universal Kahler Modulus in Warped Compactifications},''
  \href{http://dx.doi.org/10.1088/1126-6708/2009/01/036}{{\em JHEP} {\bfseries
  01} (2009) 036},
\href{http://arxiv.org/abs/0810.5768}{{\ttfamily arXiv:0810.5768 [hep-th]}}.

\bibitem{Aparicio:2015psl}
{\sc L.~Aparicio}, {\sc F.~Quevedo}, and {\sc R.~Valandro}, ``{Moduli
  Stabilisation with Nilpotent Goldstino: Vacuum Structure and SUSY
  Breaking},'' \href{http://dx.doi.org/10.1007/JHEP03(2016)036}{{\em JHEP}
  {\bfseries 03} (2016) 036},
\href{http://arxiv.org/abs/1511.08105}{{\ttfamily arXiv:1511.08105 [hep-th]}}.

\bibitem{Frey:2003dm}
{\sc A.~R. Frey}, {\sc M.~Lippert}, and {\sc B.~Williams}, ``{The Fall of
  stringy de Sitter},''
  \href{http://dx.doi.org/10.1103/PhysRevD.68.046008}{{\em Phys. Rev.}
  {\bfseries D68} (2003) 046008},
\href{http://arxiv.org/abs/hep-th/0305018}{{\ttfamily arXiv:hep-th/0305018
  [hep-th]}}.

\bibitem{Collinucci:2008sq}
{\sc A.~Collinucci}, {\sc M.~Kreuzer}, {\sc C.~Mayrhofer}, and {\sc N.-O.
  Walliser}, ``{Four-modulus 'Swiss Cheese' chiral models},''
  \href{http://dx.doi.org/10.1088/1126-6708/2009/07/074}{{\em JHEP} {\bfseries
  07} (2009) 074},
\href{http://arxiv.org/abs/0811.4599}{{\ttfamily arXiv:0811.4599 [hep-th]}}.

\bibitem{Anguelova:2009ht}
{\sc L.~Anguelova}, {\sc V.~Calo}, and {\sc M.~Cicoli}, ``{LARGE Volume String
  Compactifications at Finite Temperature},''
  \href{http://dx.doi.org/10.1088/1475-7516/2009/10/025}{{\em JCAP} {\bfseries
  0910} (2009) 025},
\href{http://arxiv.org/abs/0904.0051}{{\ttfamily arXiv:0904.0051 [hep-th]}}.

\bibitem{Kachru:2003sx}
{\sc S.~Kachru}, {\sc R.~Kallosh}, {\sc A.~D. Linde}, {\sc J.~M. Maldacena},
  {\sc L.~P. McAllister}, and {\sc S.~P. Trivedi}, ``{Towards inflation in
  string theory},'' \href{http://dx.doi.org/10.1088/1475-7516/2003/10/013}{{\em
  JCAP} {\bfseries 0310} (2003) 013},
\href{http://arxiv.org/abs/hep-th/0308055}{{\ttfamily arXiv:hep-th/0308055
  [hep-th]}}.

\bibitem{Conlon:2007gk}
{\sc J.~P. Conlon} and {\sc F.~Quevedo}, ``{Astrophysical and cosmological
  implications of large volume string compactifications},''
  \href{http://dx.doi.org/10.1088/1475-7516/2007/08/019}{{\em JCAP} {\bfseries
  0708} (2007) 019},
\href{http://arxiv.org/abs/0705.3460}{{\ttfamily arXiv:0705.3460 [hep-ph]}}.

\bibitem{Myers:1999ps}
{\sc R.~C. Myers}, ``{Dielectric branes},''
  \href{http://dx.doi.org/10.1088/1126-6708/1999/12/022}{{\em JHEP} {\bfseries
  12} (1999) 022},
\href{http://arxiv.org/abs/hep-th/9910053}{{\ttfamily arXiv:hep-th/9910053
  [hep-th]}}.

\bibitem{DeWolfe:2004qx}
{\sc O.~DeWolfe}, {\sc S.~Kachru}, and {\sc H.~L. Verlinde}, ``{The Giant
  inflaton},'' \href{http://dx.doi.org/10.1088/1126-6708/2004/05/017}{{\em
  JHEP} {\bfseries 05} (2004) 017},
\href{http://arxiv.org/abs/hep-th/0403123}{{\ttfamily arXiv:hep-th/0403123
  [hep-th]}}.

\bibitem{Herzog:2001xk}
{\sc C.~P. Herzog}, {\sc I.~R. Klebanov}, and {\sc P.~Ouyang}, ``{Remarks on
  the warped deformed conifold},'' in {\em {Modern Trends in String Theory: 2nd
  Lisbon School on g Theory Superstrings Lisbon, Portugal, July 13-17, 2001}}.
\newblock 2001.
\newblock
\href{http://arxiv.org/abs/hep-th/0108101}{{\ttfamily arXiv:hep-th/0108101
  [hep-th]}}.
\newblock

\bibitem{Klebanov:2000nc}
{\sc I.~R. Klebanov} and {\sc A.~A. Tseytlin}, ``{Gravity duals of
  supersymmetric SU(N) x SU(N+M) gauge theories},''
  \href{http://dx.doi.org/10.1016/S0550-3213(00)00206-6}{{\em Nucl. Phys.}
  {\bfseries B578} (2000) 123--138},
\href{http://arxiv.org/abs/hep-th/0002159}{{\ttfamily arXiv:hep-th/0002159
  [hep-th]}}.

\bibitem{Baumann:2007ah}
{\sc D.~Baumann}, {\sc A.~Dymarsky}, {\sc I.~R. Klebanov}, and {\sc
  L.~McAllister}, ``{Towards an Explicit Model of D-brane Inflation},''
  \href{http://dx.doi.org/10.1088/1475-7516/2008/01/024}{{\em JCAP} {\bfseries
  0801} (2008) 024},
\href{http://arxiv.org/abs/0706.0360}{{\ttfamily arXiv:0706.0360 [hep-th]}}.

\bibitem{Herzog:2002ih}
{\sc C.~P. Herzog}, {\sc I.~R. Klebanov}, and {\sc P.~Ouyang}, ``{D-branes on
  the conifold and N=1 gauge / gravity dualities},'' in {\em {Progress in
  string, field and particle theory: Proceedings, NATO Advanced Study
  Institute, EC Summer School, Cargese, France, June 25-July 11, 2002}},
  pp.~189--223.
\newblock 2002.
\newblock \href{http://arxiv.org/abs/hep-th/0205100}{{\ttfamily
  arXiv:hep-th/0205100 [hep-th]}}.
\newblock
[,383(2002)].

\bibitem{Klemm:1996ts}
{\sc A.~Klemm}, {\sc B.~Lian}, {\sc S.~S. Roan}, and {\sc S.-T. Yau},
  ``{Calabi-Yau fourfolds for M theory and F theory compactifications},''
  \href{http://dx.doi.org/10.1016/S0550-3213(97)00798-0}{{\em Nucl. Phys.}
  {\bfseries B518} (1998) 515--574},
\href{http://arxiv.org/abs/hep-th/9701023}{{\ttfamily arXiv:hep-th/9701023
  [hep-th]}}.

\bibitem{Flauger:2009ab}
{\sc R.~Flauger}, {\sc L.~McAllister}, {\sc E.~Pajer}, {\sc A.~Westphal}, and
  {\sc G.~Xu}, ``{Oscillations in the CMB from Axion Monodromy Inflation},''
  \href{http://dx.doi.org/10.1088/1475-7516/2010/06/009}{{\em JCAP} {\bfseries
  1006} (2010) 009},
\href{http://arxiv.org/abs/0907.2916}{{\ttfamily arXiv:0907.2916 [hep-th]}}.

\bibitem{Baumann:2006th}
{\sc D.~Baumann}, {\sc A.~Dymarsky}, {\sc I.~R. Klebanov}, {\sc J.~M.
  Maldacena}, {\sc L.~P. McAllister}, and {\sc A.~Murugan}, ``{On D3-brane
  Potentials in Compactifications with Fluxes and Wrapped D-branes},''
  \href{http://dx.doi.org/10.1088/1126-6708/2006/11/031}{{\em JHEP} {\bfseries
  11} (2006) 031},
\href{http://arxiv.org/abs/hep-th/0607050}{{\ttfamily arXiv:hep-th/0607050
  [hep-th]}}.

\bibitem{Krishnan:2008gx}
{\sc C.~Krishnan} and {\sc S.~Kuperstein}, ``{The Mesonic Branch of the
  Deformed Conifold},''
  \href{http://dx.doi.org/10.1088/1126-6708/2008/05/072}{{\em JHEP} {\bfseries
  05} (2008) 072},
\href{http://arxiv.org/abs/0802.3674}{{\ttfamily arXiv:0802.3674 [hep-th]}}.

\bibitem{McGuirk:2009xx}
{\sc P.~McGuirk}, {\sc G.~Shiu}, and {\sc Y.~Sumitomo}, ``{Non-supersymmetric
  infrared perturbations to the warped deformed conifold},''
  \href{http://dx.doi.org/10.1016/j.nuclphysb.2010.09.008}{{\em Nucl. Phys.}
  {\bfseries B842} (2011) 383--413},
\href{http://arxiv.org/abs/0910.4581}{{\ttfamily arXiv:0910.4581 [hep-th]}}.

\bibitem{Cohen-Maldonado:2015ssa}
{\sc D.~Cohen-Maldonado}, {\sc J.~Diaz}, {\sc T.~van Riet}, and {\sc
  B.~Vercnocke}, ``{Observations on fluxes near anti-branes},''
  \href{http://dx.doi.org/10.1007/JHEP01(2016)126}{{\em JHEP} {\bfseries 01}
  (2016) 126},
\href{http://arxiv.org/abs/1507.01022}{{\ttfamily arXiv:1507.01022 [hep-th]}}.

\bibitem{Gautason:2013zw}
{\sc F.~F. Gautason}, {\sc D.~Junghans}, and {\sc M.~Zagermann},
  ``{Cosmological Constant, Near Brane Behavior and Singularities},''
  \href{http://dx.doi.org/10.1007/JHEP09(2013)123}{{\em JHEP} {\bfseries 09}
  (2013) 123},
\href{http://arxiv.org/abs/1301.5647}{{\ttfamily arXiv:1301.5647 [hep-th]}}.

\bibitem{Blaback:2014tfa}
{\sc J.~Bl{\aa}b{\"a}ck}, {\sc U.~H. Danielsson}, {\sc D.~Junghans}, {\sc
  T.~Van~Riet}, and {\sc S.~C. Vargas}, ``{Localised anti-branes in non-compact
  throats at zero and finite $T$},''
  \href{http://dx.doi.org/10.1007/JHEP02(2015)018}{{\em JHEP} {\bfseries 02}
  (2015) 018},
\href{http://arxiv.org/abs/1409.0534}{{\ttfamily arXiv:1409.0534 [hep-th]}}.

\bibitem{Michel:2014lva}
{\sc B.~Michel}, {\sc E.~Mintun}, {\sc J.~Polchinski}, {\sc A.~Puhm}, and {\sc
  P.~Saad}, ``{Remarks on brane and antibrane dynamics},''
  \href{http://dx.doi.org/10.1007/JHEP09(2015)021}{{\em JHEP} {\bfseries 09}
  (2015) 021},
\href{http://arxiv.org/abs/1412.5702}{{\ttfamily arXiv:1412.5702 [hep-th]}}.

\bibitem{Hartnett:2015oda}
{\sc G.~S. Hartnett}, ``{Localised Anti-Branes in Flux Backgrounds},''
  \href{http://dx.doi.org/10.1007/JHEP06(2015)007}{{\em JHEP} {\bfseries 06}
  (2015) 007},
\href{http://arxiv.org/abs/1501.06568}{{\ttfamily arXiv:1501.06568 [hep-th]}}.

\bibitem{Bertolini:2015hua}
{\sc M.~Bertolini}, {\sc D.~Musso}, {\sc I.~Papadimitriou}, and {\sc H.~Raj},
  ``{A goldstino at the bottom of the cascade},''
  \href{http://dx.doi.org/10.1007/JHEP11(2015)184}{{\em JHEP} {\bfseries 11}
  (2015) 184},
\href{http://arxiv.org/abs/1509.03594}{{\ttfamily arXiv:1509.03594 [hep-th]}}.

\end{thebibliography}\endgroup

\end{document}